\documentclass{article}
\usepackage{amsmath}
\usepackage{amsfonts}
\usepackage{graphicx}
\usepackage{color}
\usepackage{subfigure}
\usepackage{amssymb,amsthm}
\usepackage{float}
\usepackage{dsfont}
\usepackage{psfrag}
\usepackage{algpseudocode}
\usepackage[margin=1.0in]{geometry}

\def\1{\mathbf{1}}

\def\L{{\mathbf L}}
\def\v{{\mathbf v}}

\def\W{{\mathbf W}}

\def\u{{\bf u}}

\newtheorem{theorem}{Theorem}[section]

\newtheorem{proposition}[theorem]{Proposition}

\floatstyle{ruled}
\newfloat{Algorithm}{tbp}{lop}[section]

\graphicspath{{Figs/}}

\title{Scalable Approach to Uncertainty Quantification and Robust Design of Interconnected Dynamical Systems}

\author{\begin{tabular}{llll}
         Andrzej Banaszuk\thanks{United Technologies Research Center, East Hartford, CT 06118, USA} & Vladimir A. Fonoberov\thanks{AIMdyn Inc., Santa Barbara, CA 93101, USA} &  Thomas A. Frewen{$^*$} & Marin Kobilarov\thanks{California Institute of Technology, Pasadena, CA 91125, USA}\\
         George Mathew\thanks{United Technologies Research Center Inc, Berkeley, CA 94705, USA} & Igor Mezic\thanks{University of California, Santa Barbara, CA 93106,USA} &  Alessandro Pinto$^\S$& Tuhin Sahai{$^*$}\\
         Harshad Sane{$^*$} & Alberto Speranzon{$^*$} & Amit Surana{$^*$} &
        \end{tabular}}
\begin{document}

\maketitle

%\address[UTRC]{}
%\address[AimDyn]{}
%\address[Caltech]{}
%\address[UTRCB]}
%\address[UCSB]{}
%%\cortext[cor1]{Corresponding Author: Email address: BanaszA@utrc.utc.com}

\begin{abstract}
%% Text of abstract
Development of robust dynamical systems and networks such as autonomous aircraft systems capable of accomplishing complex missions faces challenges due to the dynamically evolving uncertainties coming from model uncertainties, necessity to operate in a hostile cluttered urban environment, and the distributed and dynamic nature of the communication and computation resources. Model-based robust design is difficult because of the complexity of the hybrid dynamic models including continuous vehicle dynamics, the discrete models of computations and communications, and the size of the problem. We will overview recent advances in methodology and tools to model, analyze, and design robust autonomous aerospace systems operating in uncertain environment, with stress on efficient uncertainty quantification and robust design using the case studies of the mission including model-based target tracking and search, and trajectory planning in uncertain urban environment. To show that the methodology is generally applicable to uncertain dynamical systems, we will also show examples of application of the new methods to efficient uncertainty quantification of energy usage in buildings, and stability assessment of interconnected power networks.
\end{abstract}

\vspace*{1cm}
\noindent\textbf{Keywords}\\
\noindent Aerospace Systems; Graph Decomposition; Model Based Design;  Motion Planning; Uncertainty
\vspace*{1cm}

% \linenumbers

%% main text
\section{Introduction}\label{sec:Intro}
In this paper we discuss methodology and tools enabling model-based design of complex interconnected dynamical systems that would operate as desired in the presence of uncertainty in models and environment. A particular challenge problem we are going to consider is design of unmanned aerospace systems that can conduct complex missions in a safe, predictable, and robust manner.  Model-based robust design is difficult for such systems because of the environmental and modeling uncertainties, and the overall system complexity. Dynamically evolving uncertainties arise from model uncertainties, necessity to operate in a adversarial  cluttered urban environment, and the distributed and dynamic nature of the communication and computational resources. Complexity results from the hybrid dynamic models including continuous vehicle dynamics, the discrete models of computations and communications, and the size of the problem. We overview recent advances in methodology and tools for managing uncertainty and complexity in such interconnected dynamical systems. We illustrate this methodology for modeling, analysis and robust design of autonomous aerospace systems using case studies, which include low level autonomous flight in urban environment and model-based target tracking and search. To show the generality of our approach, we also include examples of efficient uncertainty quantification of energy usage in buildings, and stability assessment of interconnected power networks.
%Development of autonomous aerospace systems capable of accomplishing complex missions faces challenges due to the dynamically evolving uncertainties coming from model uncertainties, necessity to operate in a hostile cluttered
%urban environment, and the distributed and dynamic nature
%of the communication and computation resources. Model-based robust design is difficult
%because of the complexity of the hybrid dynamic models including continuous vehicle dynamics,
%the discrete models of computations and communications, and the size of the problem. We will overview recent advances in methodology and tools to model, analyze, and design robust autonomous aerospace systems operating in uncertain environment, with stress on efficient uncertainty quantification and robust design using the case studies of the missions including low level autonomous flight in urban environment and model-based target tracking and search. To show that the methodology is generally applicable to uncertain dynamical systems, we will also show examples of application of the new methods to efficient uncertainty quantification of energy usage in buildings, and stability assessment of interconnected power networks.

The objective of Robust Uncertainty Management (RUM) project \cite{RUM_final_report} was to develop methodology and tools for quantifying uncertainty in ways that are orders of magnitude faster than Monte Carlo with near-linear scaling in the system size, and demonstrate them in molecular dynamics and UAV search challenge problems. Several Uncertainty Quantification (UQ) methods including Polynomial Chaos (PC), and new Stochastic Response Surface, and Dynamic Sampling methods were applied to calculate the mean and variance of the phase transition temperature in molecular dynamics calculations with 10000 atoms 2000 times faster than using Monte Carlo sampling. The new search strategies described in \cite{RUM_final_report} achieved 2x reduction in the median search time compared with a standard lawnmower approach in a challenge problem including 50 simulated UAVs searching for a stationary target in a complex terrain using noisy sensors with uncertain footprint.  One method used a trajectory planner based on the use of a library of pre-computed elementary UAV trajectory segments that individually satisfy the vehicle dynamics and can be interlocked to produce large scale roadmaps. In this paper we provide an overview of selected methods and tools for Uncertainty Management and Stochastic Design Methodology that started within the RUM project and were continued under the Autonomous and Intelligent Systems initiative at United Technologies Research Center (UTRC), including strong collaboration with researchers at UCSB, Caltech, and Aimdyn, Inc.

The key idea that enables scalable computations in interconnected dynamic networks presented in Section~\ref{Bottom up graph decomposition} of this paper is  that of {\em decomposing} large networks of dynamical components into {\em weakly connected subnetworks} using spectral properties of a graph Laplacian constructed from a Jacobian matrix of the underlying set of ODEs describing the dynamic network. This graph decomposition method for dynamic networks was first introduced in \cite{Igor}. The decomposition allows to parallelize the problem of solving large ODEs using the method called Waveform Relaxation \cite{wave}. Iterations across weak connections yield a provably convergent method under the assumption that underlying set of ODEs satisfies the Lipshitz condition. A new efficient and easily parallelizable method of obtaining spectral decomposition applicable to large dynamic networks has been introduced in \cite{WaveEqnClus} and is described in Section~\ref{Bottom up graph decomposition}. Rather than relying on random walk methods used in state of the art methods for distributed computations of spectral properties of matrices, the method relies on using the Graph Laplacian corresponding to the dynamic network of interest to define and solve a wave equation.

In Section~\ref{Scalable Uncertainty Quantification} we show how the graph decomposition introduced in Section~\ref{Bottom up graph decomposition}  can be generalized to scalable Uncertainty Quantification using a method called Probabilistic Waveform Relaxation introduced in  \cite{pwr_cdc10}. Once the dynamic network is decomposed into weakly connected subnetworks, efficient methods for Uncertainty Quantification are applied to subnetworks in an iterative manner. The method is called Probabilistic Waveform Relaxation and is amenable to parallel computations. We demonstrate the method on examples from energy flow in buildings and power networks.

In Section~\ref{sec:algorithm} we address another aspect of Uncertainty Management: {\em design} of search and tracking algorithms for unmanned aerospace systems that allow to reduce the uncertainty of location of stationary or mobile targets. For search and tracking applications, it is important to design uniform coverage dynamics for mobile sensors. A uniform coverage control algorithm would ensure that the amount of time spent by the sensors observing a neighborhood is proportional to the probability of finding a target in the neighborhood. For the search of a stationary target, the uncertainty in the position of the target can be specified in terms of a fixed probability distribution. The Spectral Multiscale Coverage algorithm proposed in (\cite{smc_cdc09} and \cite{smc_physd09}), makes the sensors move so that points on the sensor trajectories uniformly sample this stationary probability distribution. Uniform coverage dynamics coupled with sensor observations helps to reduce the uncertainty in the position of the target. In \cite{MAS}, it has been demonstrated that in the presence of various uncertainties, uniform coverage based search strategies outperform lawnmower-type search strategies.  In a recent paper \cite{dsmc_cdc10} we extended the SMC algorithm to the case of moving targets.

While in Section~\ref{sec:algorithm} we consider a simplistic model of a mobile sensor and no obstacles, in Section~\ref{sec:Robust Path Planning} we consider a considerably more difficult problem of optimizing a motion of a vehicle described by a realistic model of motion in an obstacle-rich environment. The vehicle is subject to constraints arising from underactuated dynamics, actuator bounds, and obstacles in the environment.
We assume that the vehicle is equipped with a sensor measuring the relative positions of obstacles. The problem has no closed-form solution since both the dynamics and constraints are nonlinear. Gradient-based optimization is not suitable unless a good starting guess is chosen since the obstacles  impose many local minima. In this paper we also employ a graph-based search but unlike in standard discrete search, the nodes of the tree are sampled from the original continuous space and the edges correspond to
trajectories satisfying any given dynamics and general constraints.
Our approach is based on a recent methodology under active development in the robotics community known as
sampling-based motion planning which includes the
\emph{rapidly-exploring random tree} (RRT)~\cite{La2006} and the
\emph{probabilistic roadmap} (PRM)~\cite{ChLyHuKaBuKaTh2005}. In
our framework  samples are connected through sequences of locally
optimized precomputed motion primitives. The two main advantages of the approach
is that the locally optimal motion primitives used are computed offline and
that standard graph search methods can be used for finding global solutions efficiently. Note that even if the model of the vehicle is assumed deterministic and perfectly known, the
introduction of sampling-based motion planning introduces uncertainty in the trajectory planing process. In particular, one can only characterize the probability of computing a feasible path as a function of the number of samples.

In Section~\ref{sec:Design flows} we show how the techniques presented in this paper are organized into a design process. We shows how analysis and synthesis tools can be effectively used in a design flow that is based on the notion of abstraction levels. While Section 2 and 3 present decomposition techniques that apply to one particular abstraction level, the methodology in Section 6 shows how a design problem is vertically decomposed into refinement steps. For example, search and tracking algorithms can rely on an abstraction of the low level path planner. In this section we also discuss the challenges in the integration and deployment of tools to serve development of industrial applications. We conclude the section by showing our efforts in this direction.

\section{Bottom up graph decomposition}\label{Bottom up graph decomposition}

In recent years, there has been an exponential increase of interest in large interconnected
systems, such as sensors networks, thermal networks, social networks, internet,
biochemical networks, power networks, communication networks etc. These systems are
characterized by complex behavior arising because of interacting
subsystems. Often these interactions are weak which can be exploited to simplify and accelerate the simulation, analysis and design of such systems by suitably decomposing them. To facilitate this decomposition it is convenient to model such networked systems by a graph of interacting subsystems. Consequently, graph theoretic methods have been recently applied and extended to study these
systems. To illustrate this idea, consider a large interconnected system described by a system of differential equation
\begin{eqnarray}
  \dot{x_1}&=&f_1(\mathbf{x},\mathbf{\xi}_1,t),\notag\\
    \vdots\notag\\
  \dot{x_n}&=&f_n(\mathbf{x},\mathbf{\xi}_n,t),\label{complexsys}
\end{eqnarray}
where, $\mathbf{f}=(f_1,f_2,\cdots,f_n)\in\mathbb{R}^n$ is a smooth vector field, $\mathbf{x}=(x_1,x_2,\cdots,x_n)\in \mathbb{R}^n$ are state variables, $\mathbf{\xi}_i\in R^{p_i}$ is vector of (possibly uncertain) parameters affecting the $i-$th system. Let $\mathbf{\xi}=(\mathbf{\xi}_1^T,\cdots,\mathbf{\xi}_n^T)^T\in \mathbb{R}^p$ be the $p=\sum_{i=1}^n p_i$ dimensional vector of parameters affecting the complete system. The solution to initial value problem $\mathbf{x}(t_0)=\mathbf{x}_0$ will be denoted by $\mathbf{x}(t;\mathbf{\xi})$, where for brevity we have suppressed the dependence of solution on initial time $t_0$ and initial condition $\mathbf{x}_0$. Given the set of state variables $x_1,\cdots,x_n$ and some notion of dependence $\W_{ij}\ge 0,i=1,\cdots,n,j=1,\cdots,n$ between pairs of state variables, a graph can be constructed. The vertices in this graph represent the states variables $x_i$ and two vertices are connected with an edge of weight $\W_{ij}$.  In order to quantify coupling strength $\W_{ij}$ between nodes or state variables, one could use for example \cite{igorcdc}
\begin{equation}\label{similiarity}
\W_{ij}=\frac{1}{2}[|\overline{J}_{ij}|+|\overline{J}_{ji}|],
\end{equation}
where, $\overline{J}=[\frac{1}{T_s}\int_{t_0}^{t_0+T_s}J_{ij}(\mathbf{x}(t;\mathbf{\xi}_m),\mathbf{\xi}_m,t)dt]$,
is time average of the Jacobian,
\begin{equation}\label{Jac}
J(\mathbf{x},\mathbf{\xi},t)=\left[\frac{\partial f_i(\mathbf{x}(t;\mathbf{\xi}),\mathbf{\xi},t)}{\partial x_j}\right],
\end{equation}
computed along the solution $\mathbf{x}(t;\mathbf{\xi})$ of the system (\ref{complexsys}) for nominal value of parameters $\mathbf{\xi}_m$. The decomposition or clustering problem can now be formulated as follows: find a partition of the graph such that the edges between different components have a very low weight and the edges within a component have high weights. The main tool for accomplishing such a decomposition is the graph Laplacian. In particular, spectral properties of the Laplacian
matrix~$L$ associated to such graphs provide very useful
information for the analysis as well as the design of
interconnected systems. The computation of eigenvectors of the
graph Laplacian~$L$ is the cornerstone of spectral graph
theory~\cite{Chung,Tutorial}, and it is well known that the sign
of the second (and successive eigenvectors) can be used to
cluster graphs~\cite{Fiedler,Fiedler2}.

Let $\mathcal{G}=(V,E)$ be a graph with vertex set~$V =
\{1,\dots,N\}$ and edge set $E\subseteq V\times V$, where a
weight~$\W_{ij}\in \mathds{R}$ is associated with each edge
$(i,j)\in E$, and $\W \in \mathds{R}^{N\times N}$ is the
weighted adjacency matrix of~$\mathcal{G}$. We assume that
$\W_{ij}=0$ if and only if $(i,j) \notin E$. The (normalized)
graph Laplacian is defined as,
\begin{align*}
    \L_{ij} = \begin{cases}
                1 & \mbox{if}\: i = j\\
                -\W_{ij}/\sum_{\ell=1}^N \W_{i\ell} & \mbox{if}\: (i,j) \in E\\
                0   & \mbox{otherwise}\,.
             \end{cases}
\end{align*}

In this work we only consider undirected graphs. The smallest
eigenvalue of the Laplacian matrix is $\lambda_1 = 0$, with an
associated eigenvector $\v^{(1)}=\1=\left[1,1,\dots,1\right]^T$.
Eigenvalues of~$\L$ can be ordered as, $ 0 = \lambda_1 \leq
\lambda_2 \leq \lambda_3 \leq \cdots \leq \lambda_N$ with
associated eigenvectors $\1, \v^{(2)}, \v^{(3)}\cdots
\v^{(N)}$~\cite{Tutorial}. It is well known that the
multiplicity of~$\lambda_1$ is the number of connected
components in the graph~\cite{Mohar}. We assume in the following
that $\lambda_1<\lambda_2$ (the graph does not have trivial
clusters). We also assume that there exists unique cuts that
divide the graph into $k$ clusters. In other words, we assume
that there exist $k$ distinct eigenvalues~\cite{UlrikeDist}.

Given the Laplacian matrix~$\L$, associated with a
graph~$\mathcal{G}= (V,E)$, spectral clustering divides
$\mathcal{G}$ into two clusters by computing the sign of the~$N$
elements of the second eigenvector~$\v^{(2)}$, or Fiedler
vector~\cite{Fiedler2,Tutorial}.

As the clustering is decided by the eigenvectors/eigenvalues of
the Laplacian matrix one can use standard matrix algorithms for
such computation~\cite{GolubVanLoan96}. Since we are considering
multi-agent systems, the execution of these standard algorithms
is infeasible in decentralized settings. To address this issue,
algorithms for distributed eigenvector computations have been
proposed~\cite{KempeMcSherry08}. These algorithms, however, are
also (like the algorithm in~\cite{Spielman}) based on the slow
process of performing random walks on graphs.

The slowest step in the distributed computation of eigenvectors
is the simulation of a random walk on the graph. This procedure
is equivalent to solving the discrete version of the heat
equation on the graph. The connection between spectral
clustering and the heat equation was also pointed out
in~\cite{Raphy1,Raphy2}.

In a theme similar to M. Kac's question ``Can one hear the shape
of a drum?''~\cite{DrumShape}, we demonstrate that by simulating
the wave equation over a graph, nodes can ``hear'' clusters of
the graph Laplacian using only local information. Moreover, we
demonstrate, both theoretically and on examples, that the wave
equation based algorithm is orders of magnitude faster than
random walk based approaches for graphs with large mixing times.
The overall idea of the wave equation based approach is to
simulate, in a distributed fashion, the propagation of a wave
over a graph and capture the frequencies at which the graph
``resonates''. In this paper we show that from such frequencies
it is possible to exactly recover the eigenvectors of~$L$ from
which the clustering problem is solved. We also provide
conditions the wave must satisfy in order to cluster a graph
using the proposed method.

\subsection{Wave Equation Based Clustering}
Similar to the case of the heat equation, the solution of the
wave equation
\begin{equation}
    \frac{\partial^{2} u}{\partial t^{2}}=c^{2}\Delta u\,.
    \label{Waveeqn}
\end{equation}
can be expanded in terms of the eigenvectors of
the Laplacian. However, unlike the heat equation where the
solution eventually converges to the first eigenvector of the
Laplacian, in the wave equation all the eigenvectors remain
eternally excited~\cite{Evans}. Here we use the discretized wave
equation on graphs  to develop a simple, yet powerful, eigenvector computation algorithm. The main steps of the algorithm are shown as Algorithm~\ref{alg:WaveAlg}. Note that some
properties of the wave equation on graphs have been studied
in~\cite{WaveGraphProp}.
%\begin{align}
%    \u_{i}(t) = 2\u_{i}(t-1) - \u_{i}(t-2) - c^2\displaystyle\sum_{j\in\mathcal{N}(i)}\L_{ij}
%    \u_{j}(t-1)\,.
%    \label{onenodewave}
%\end{align}
%Note that at each node (node~$i$ in
%the algorithm) one only needs nearest neighbor weights~$\L_{ij}$
%and the scalar quantities $\u_{j}(t-1)$ also at nearest
%neighbors. We emphasize, again, that $\u_{i}(t)$ is a scalar
%quantity and \texttt{Random}($[0,1]$) is a random initial
%condition on the interval $[0,1]$. .

\begin{Algorithm}
    \caption{Wave equation based eigenvector computation algorithm for node~$i$. At node~$i$ one computes the sign of the~$i$-th component of the first~$k$ eigenvectors. The vector~$\v^{(j)}_{i}$ is the $i$-th component of the $j$-th eigenvector. The cluster number is obtained from interpreting the vector of $k$ signs as a binary number.}\label{alg:WaveAlg}
    \begin{algorithmic}[1]
    \State $\u_{i}(0) \leftarrow \texttt{Random}\:([0,1])$
    \State $\u_{i}(-1) \leftarrow \u_{i}(0)$
    \State $t\leftarrow 1$
    \While{$t < T_{max}$}
         \State \begin{tabular}{l}
         $\u_{i}(t) \leftarrow 2\u_{i}(t-1)-\u_{i}(t-2) -$\\
         $\qquad \qquad c^2 \sum_{j\in\mathcal{N}(i)}\L_{ij}\u_{j}(t-1)$
         \end{tabular}
         \State $t\leftarrow t+1$
    \EndWhile
    \State $Y\leftarrow \texttt{FFT}\:(\left[\u_{i}(1),\dots\dots,\u_{i}(T_{max})\right])$
    \For{$j \in \{1,\dots,k\}$}
        \State $\theta_{j} \leftarrow \texttt{FrequencyPeak}\:(Y,j)$
        \State $\v^{(j)}_{i} \leftarrow \texttt{Coefficient}(\theta_{j})$ %\Comment {either one of $\sin(t\theta_{j})$ or $\cos(t\theta_{j})$}
        \If {$\v^{(j)}_{i} > 0$}
            \State $C_j \leftarrow 1$
        \Else
            \State $C_j \leftarrow 0$
        \EndIf
        \EndFor
    \State ClusterNumber $\leftarrow \sum_{j=1}^k C_j 2^{j-1}$
\end{algorithmic}
\end{Algorithm}

\begin{proposition}
The clusters of the graph~$\mathcal{G}$, as determined by the
sign of the elements of Fiedler's eigenvector (as well as higher
order eigenvectors) of~$\L$, can be equivalently computed using
the frequencies and coefficients obtained from the Fast Fourier
Transform of $(\u_i(1),\dots,\u_i(T))$, for all~$i$. Here~$\u_i$
is governed by the wave equation on the graph with initial
condition~$\u(-1)=\u(0)$.
\end{proposition}
\begin{proposition}
The wave equation iteration (Step 5 in Algorithm ~\ref{alg:WaveAlg}) is stable on any
graph if the speed of the wave satisfies the following
inequality, $ c \leq \sqrt{2}$.
\end{proposition}
%\begin{proof}
Proof of these propositions can be found in ~\cite{WaveEqnClus}.
%\end{proof}

%\begin{remark}
%Note that if the matrix~$L$ has distinct eigenvalues, then the
%wave equation can also be used as distributed algorithm for
%eigenvector and eigenvalue computation.
%\end{remark}
%
%\begin{remark}
%Our algorithm is also attractive from a communication point of
%view. In~\cite{KempeMcSherry08} entire matrices need to be
%passed from one node to another. In our algorithm only scalar
%quantities $\u_{j}$ need to be communicated.
%\end{remark}

%where $\lambda_N$ is the largest eigenvalue of the Laplacian.
%\end{proposition}
%\begin{proof}
%See~\cite{WaveEqnClus}.
%\end{proof}

An important quantity related to the wave equation based
algorithm is the time needed to compute the eigenvalues and
eigenvectors components. The distributed eigenvector algorithm
proposed in~\cite{KempeMcSherry08} converges at a rate of
$O(\tau\log^{2}(N))$ , where~$\tau$ is the mixing time of the
Markov chain associated with the random walk on the graph. We
derive a similar convergence bound for the wave equation based
algorithm.

It is evident that one needs to resolve the lowest frequency to
cluster the graph~\cite{WaveEqnClus}. Let us assume that one
needs to wait for~$\eta$ cycles of the lowest frequency to
resolve it successfully (i.e. the number of cycles needed for a
peak to appear in the FFT)\footnote{The constant~$\eta$ is
related to the FFT algorithm and independent of the graph.
Typically 6-7 cycles of the lowest frequency are necessary to
discriminate it.}. The time needed to cluster the graph based on
the wave equation is,
\begin{equation}
    T = O\left(\frac{1}{\arccos(e^{-1/\tau})}\right) +
    O(N)\,. \label{eq:Waveconvfinal}
\end{equation}
For details of the derivation see~\cite{WaveEqnClus}.

\begin{figure}[t!]
  \centering
  \includegraphics[width=0.5\hsize]{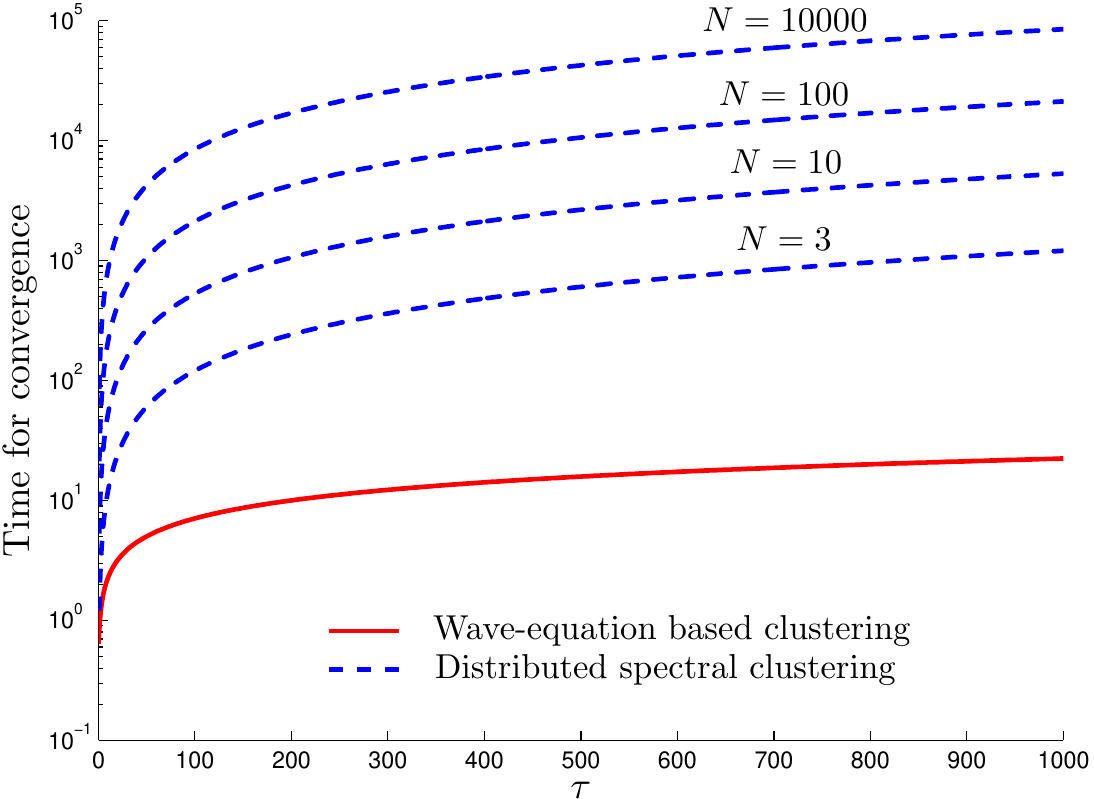}\\
  \caption{Comparison of convergence between distributed algorithm~\cite{KempeMcSherry08} and proposed wave equation
  algorithm. Wave equation based algorithm has better scaling with~$\tau$ for graphs of any
  size (given by $N$). Of course the plots are upper bounds on the convergence speed, but they are anyway indicative.}\label{ConvComp}
\end{figure}

\subsection{Numerical Examples}
To check the accuracy of our distributed wave equation based
clustering algorithm, we demonstrate our approach on several examples.

\subsubsection{Social Network}
\begin{figure}[t!]
    \begin{center}
        \includegraphics[width=0.55\hsize]{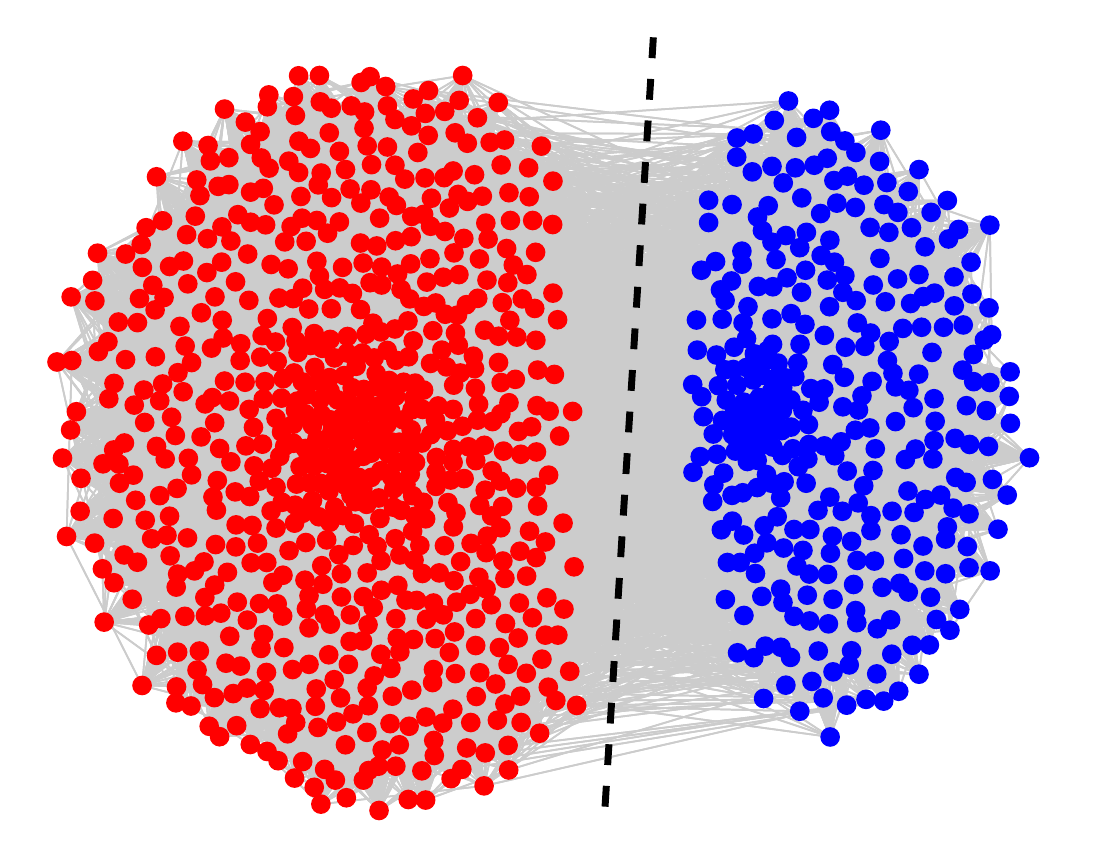}
    \end{center}
    \caption{\label{fig:Fortunato}A Fortunato community detection benchmark with $1000$ nodes and $99084$ edges. Wave equation based clustering computes the graph cut exactly.}
\end{figure}

The first example is the Fortunato benchmark graph~\cite{Fortunato} with $1000$ nodes
and $99084$ edges. Here the graph represents a social network
with two communities (clusters) with $680$ and $320$ nodes
respectively. Since these clusters are known apriori we can
check the accuracy of our approach. We find that the wave
equation based clustering computes the graph cut exactly. These
clusters are shown in Fig.~\ref{fig:Fortunato}.

\subsubsection{Building Thermal Network} \label{sec:buildeg}
\begin{figure}[th!]
    \begin{center}
    \subfigure[]{\includegraphics[scale=0.4]{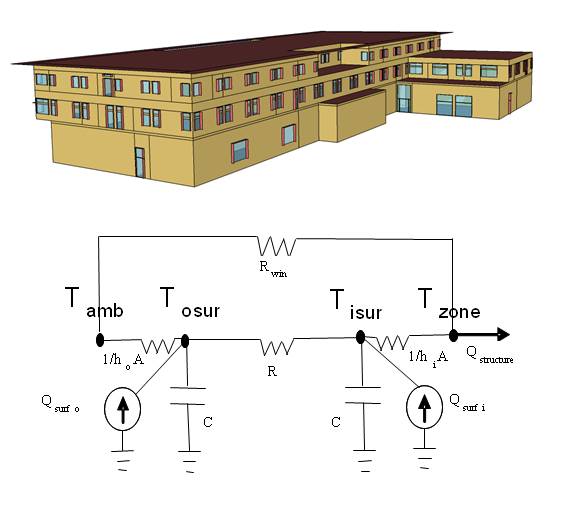}}\hfill
    \subfigure[]{\includegraphics[scale=0.4]{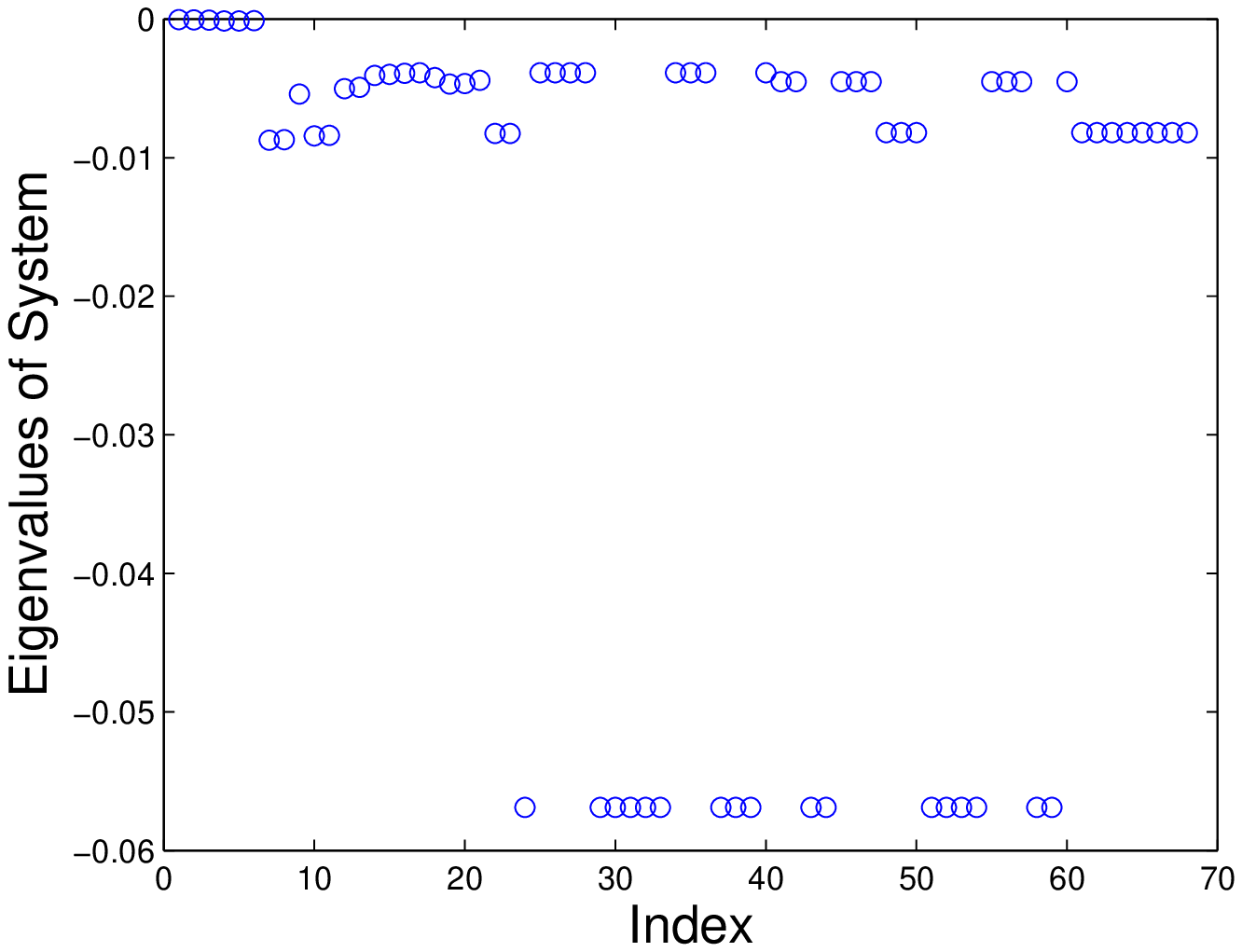}}\hfill
    \subfigure[]{\includegraphics[scale=0.4]{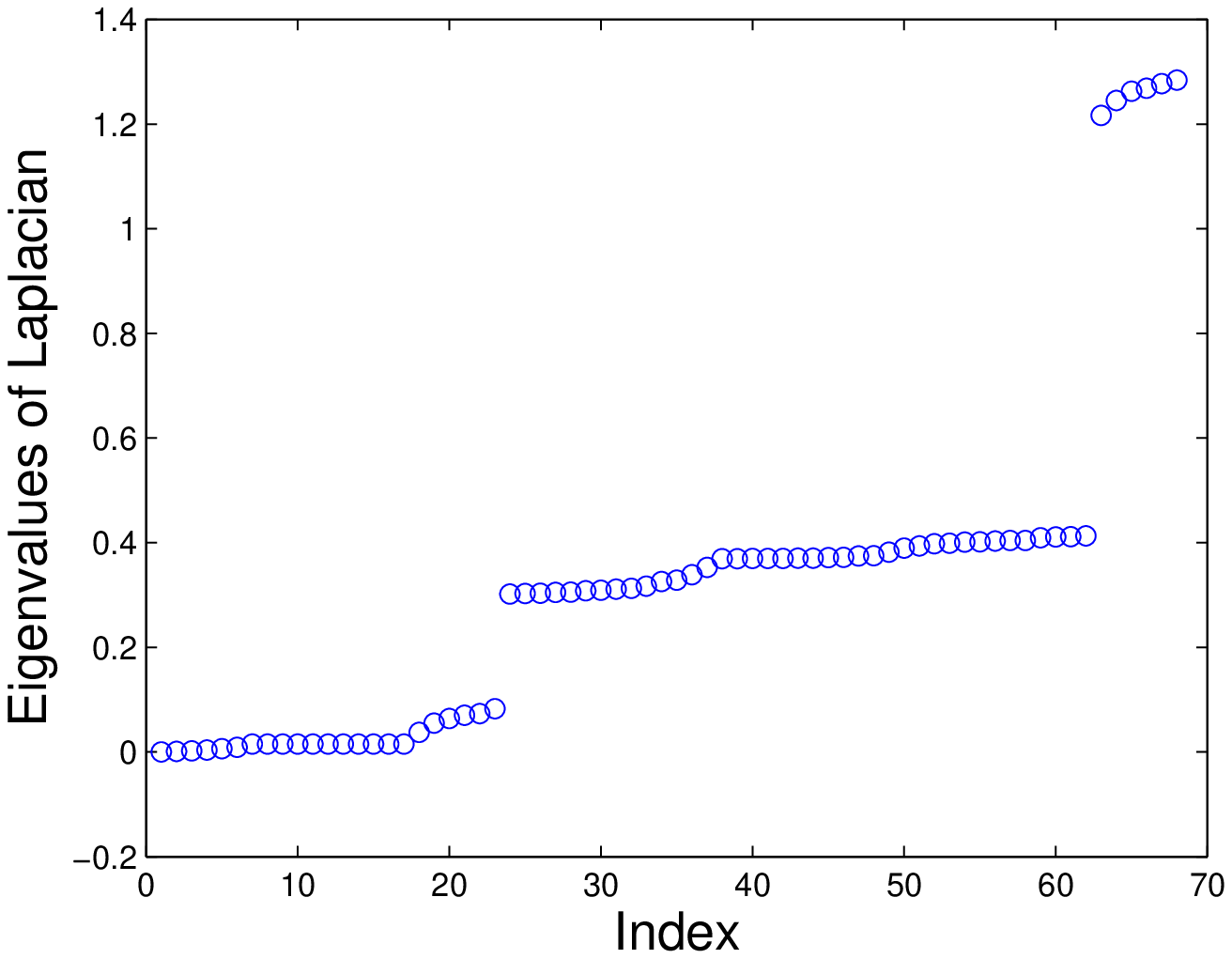}}\hfill
    \end{center}
    \caption{(a) Schematic of a building and thermal network model.(b) Shows the there bands of eigenvalues of the time averaged $A(t;\xi)$ for nominal parameter values. (c) First spectral gap in graph Laplacian revealing $23$ subsystems in the network model.} \label{Fig:spectraldecomposition}
\end{figure}

For energy consumption computation, a building can be represented in terms of a reduced order thermal network model \cite{zheng2009},
\begin{equation}\label{Buildmodel}
\frac{dT}{dt}=A(t;\xi)T+B\left(
                       \begin{array}{c}
                         Q_{e}(t) \\
                         Q_{i}(t) \\
                       \end{array}
                     \right)
\end{equation}
where, $T\in R^n$ is a vector comprising of internal zone air temperatures, and internal and external wall temperatures; $A(t;\xi)$ is the time dependent matrix with $\xi$ being parameters, and $Q_e$ and $Q_i$ represent the solar radiation and occupant load. Figure \ref{Fig:spectraldecomposition}a shows schematic of a building and key elements of a thermal network model. For the particular building considered, the network model is derived by lumping one of the building floors into 6 zones and comprises of $68$ state variables. This model admits a decomposition into $23$ subsystems, as revealed by the spectral graph approach (see figure \ref{Fig:spectraldecomposition}c). This decomposition is consistent with three different time scales (associated with external and internal wall temperature, and internal zone temperatures) present in the system, as shown by the three bands in figure \ref{Fig:spectraldecomposition}b.

\subsection{Extensions}
Current work includes the extension of the wave equation based
algorithm for dynamic networks. This is clearly very important
in situations where the weights on the edges of the graph are
time varying. The authors are also extending these algorithms to
cluster multi-attributed graphs (with multiple edges between any
two nodes) and pursuing multi-scale versions of the
wave-equation based clustering method. We are also developing methods which exploit the system decomposition to accelerate simulation and analysis of networked dynamical systems. On such application to uncertainty quantification is described in the next section.

\section{Scalable Uncertainty Quantification}\label{Scalable Uncertainty Quantification}
Uncertainty Quantification (UQ) methods provide means of calculating probability distribution of system outputs, given probability distribution of input parameters. Outputs of interest could include for example, latency in communication network, power quality and stability of power networks, and energy usage in thermal networks. The standard UQ methods such as Monte Carlo(MC) \cite{MCS} either exhibit poor convergence rates or others such as Quasi Monte Carlo (QMC) \cite{nets}\cite{lattice}, generalized Polynomial Chaos(gPC) \cite{gPC} and the associated Probabilistic Collocation method (PCM) \cite{xiu}, suffer from the curse of dimensionality (in parameter space), and become practically infeasible when applied to network as a whole. Improving these techniques to alleviate the curse of dimensionality is an active area of current research, see \cite{cursedim} and references therein: notable methods include sparse-grid collocation method \cite{sparse},\cite{MEPCM} and  ANOVA decomposition \cite{Anova} for sensitivity analysis and dimensional reduction of the uncertain parametric space. However, none of such extensions exploits the underlying structure and dynamics of the networked systems which can often prove advantageous. For example, authors in \cite{gp1,gp2} used graph decomposition to facilitate stability and robustness analysis of large-scale interconnected dynamical systems. Mezic et al. \cite{igorcdc} used graph decomposition in conjunction with Perron Frobenius operator theory to simplify the invariant measure computation and uncertainty quantification, for a particular class of networks. While these approaches exploit the underlying structure of the system, they do not take advantage of the weakly coupled dynamics of the subsystems.

\subsection{Probabilistic Waveform Relaxation}
We propose an iterative UQ approach that exploits the weak interactions among subsystems in a networked system to overcome the dimensionality curse associated with traditional UQ methods. This approach relies on integrating graph decomposition techniques and waveform relaxation schemes with probabilistic collocation and generalized polynomial chaos. Graph decomposition to identify weakly interacting subsystems was described in section \ref{Bottom up graph decomposition}. Waveform relaxation \cite{wave}, a parallelizable iterative method, on the other hand, exploits this decomposition and evolves each subsystem forward in time independently but coupled with the other subsystems through their solutions from the previous iteration. At each waveform relaxation iteration we propose to apply PC at subsystem level and use gPC to propagate the uncertainty among the subsystems. Since UQ methods are applied to relatively simpler subsystems which typically involve a few parameters, this renders a scalable iterative approach to UQ in complex networks. We refer to this iterative UQ approach as \emph{probabilistic waveform relaxation} (PWR).

%\begin{figure}[th!]
%\begin{center}
%\includegraphics[width=6cm]{PWR}\\
%\caption{Schematic description of the PWR framework.}\label{fig0}
%\end{center}
%\end{figure}

We illustrate the proposed PWR framework through an example of parametric
uncertainty in a simple system (\ref{complexsys}). The PWR framework also applies
to DAE models and stochastic processes as will be described later. Consider the following
system:
\begin{eqnarray}
  \dot{x_1}&=&f_1(x_1,x_2,\xi_1,t),\notag\\
  \dot{x_2}&=&f_2(x_1,x_2,x_3,\xi_2,t),\label{complexsys}\\
  \dot{x_3}&=&f_3(x_2,x_3,\xi_3,t)\notag
\end{eqnarray}
Here, $\xi_i$ is a random variable affecting the $i^{th}$ system ($i=1,2,3$). To keep
our exposition streamlined, we will assume that parameters
$\xi_i$ are mutually independent, each having probability
density $w_i=w_i(\xi_i)$. We further assume that the three subsystems in
(\ref{complexsys}) weakly interact with each other, so that subsystem $2$ weakly
affects subsystem $1$ and $3$ and vice versa.
In the example we use full grid PCM as the subsystem level UQ method and gPC as
the uncertainty propagation basis \cite{pwr_cdc10}.

In standard gPC, states $x_i$ are expressed as orthogonal polynomials of the input
random parameters, and different orthogonal polynomials can be chosen to achieve
better convergence \cite{gPC}. Consider a $P$-variate space $W^{\Lambda}$,
formed over any random parameter subset $\Lambda \subseteq \Sigma=\{\xi_1,\xi_2,\xi_3\}$,
by taking tensor product of a one-dimensional orthogonal polynomial space associated
with each random variable $\xi_i, i=1,2,3$. We denote the basis elements of
$W^{\Lambda}$ by $\Psi_{j}^{\Lambda},j=1,\cdots,N_{\Lambda}$, where
$N_{\Lambda}=\mbox{dim}(W^{\Lambda})$. The state variables can be expanded
in polynomial chaos basis associated with $W^{\Sigma}$, as
\begin{equation}\label{expx}
x_i^{\Sigma}(t,\mathbf{\xi}) =
	\sum_{j=1}^{N_{\Sigma}} a_{j}^i(t)\Psi_j^{\Sigma}(\mathbf{\xi}),\quad i=1,2,3
\end{equation}
where, $a_j^i(t)$ are the modal coefficients, and the sum has been truncated to
a finite order. When applied to differential equations with random parameters,
the gPC expansion is typically combined with Galerkin projection, such that the
resulting set of equations for the expansion coefficients are deterministic and
can be solved via conventional numerical techniques.

Note that in the gPC expansion (\ref{expx}), the system states are expanded in terms of
all the random variables $\xi$ affecting the entire system. From the structure of
system (\ref{complexsys}) it is clear that the $1^{st}$ subsystem is directly
affected by the parameter $\xi_1$ and indirectly by parameter $\xi_2$ through the
the state $x_2$. We neglect second order effect of $\xi_3$ on $x_1$.
A similar statement holds true for subsystem $3$, while subsystem $2$ will be
weakly influenced by $\xi_1$ and $\xi_3$ through states $x_1$ and $x_3$, respectively.
This structure can be used to simplify the Galerkin projection as follows. For $x_1$
we consider the gPC expansion over $\Sigma_1=\Lambda_1\bigcup\Lambda_1^c$,
\begin{equation}%\label{expx}
	x_1^{\Sigma_1}(t,\xi_1,\xi_2) =
	\sum_{j=1}^{N_{\Sigma_1}} \overline{a}_{j}^1(t)\Psi_j^{\Sigma_1}(\xi_1,\xi_2),
\end{equation}
where, $\Lambda_1=\{\xi_1\}$, $\Lambda_1^c=\{\xi_2\}$.
Similarly, one can consider simplification for $x_3^{\Sigma_3}(t,\xi_1,\xi_3)$.
We also introduce the following two projections associated with the state $x_2$:
\begin{equation}\label{proj}
	P^{2,i}(x_2^{\Sigma_2}) =
	\sum_{j=1}^{N_{\Sigma_i}} \left\langle x_2^{\Sigma_2},\Psi_{j}^{\Sigma_i} \right\rangle
	\Psi_{j}^{\Sigma_i}.
\end{equation}
where $i=1,3$  and $\langle \cdot,\cdot \rangle$ is the appropriate inner product.
With these expansions, and using standard Galerkin projection we obtain
the following system of deterministic equations
\begin{equation}\label{agp}         %\label{galproj1}
	\dot{\overline{a}}^i_{j}=\overline{F}_{j}^i(\overline{\mathbf{a}},t),
	~~~j=1,\ldots , N_{\Sigma_{i}}
\end{equation}
with appropriate initial conditions, where
\begin{eqnarray}\label{Fbar}
	\overline{F}_{j}^1(\overline{\mathbf{a}}) &=& \int_{\Gamma_{\Sigma}}f_1(x_1^{\Sigma_1},P^{2,1}
	(x_2^{\Sigma_2}),\xi_1,t)\Psi_j^{\Sigma_1}(\mathbf{\xi})
	\mathbf{w}(\mathbf{\xi})d\mathbf{\xi},\notag, \\
\overline{F}_{j}^2(\overline{\mathbf{a}})&=&\int_{\Gamma_{\Sigma}}f_2(x_1^{\Sigma_1},x_2^{\Sigma_2},x_3^{\Sigma_3},\xi_2,t)\Psi_j^{\Sigma_2}(\mathbf{\xi})
\mathbf{w}(\mathbf{\xi})d\mathbf{\xi},\notag\\
\overline{F}_{j}^3(\overline{\mathbf{a}})&=&\int_{\Gamma_{\Sigma}}f_3(P^{2,3}(x_2^{\Sigma_2}),x_1^{\Sigma_1},\xi_3,t)\Psi_j^{\Sigma_3}(\mathbf{\xi})
\mathbf{w}(\mathbf{\xi})d\mathbf{\xi},\notag
\end{eqnarray}
and $\overline{\mathbf{a}} =
(\overline{\mathbf{a}}_1, \overline{\mathbf{a}}_2, \overline{\mathbf{a}}_3)^T$,
with $\mathbf{a}_i = (\overline{a}^{i}_{1},\cdots,\overline{a}^{i}_{N_{\Sigma_{i}}})$. We will refer to (\ref{agp}) as an approximate Galerkin projected system. The notion of approximate Galerkin projection in more general setting can be found in \cite{pwr_cdc10}. In many instances Galerkin projection may not be possible due to unavailability of
direct access to the system equations (\ref{complexsys}). In many other instances
intrusive methods are not feasible even in cases when the system equations are available, because of the cost of deriving
and implementing a Galerkin system within available computational tools. We next describe the non-intrusive PWR using probabilistic collocation, which does not require the Galerkin projection (\ref{agp}) explicitly.

%Similarly, let
%$\overline{\mathbf{F}} =
%(\overline{\mathbf{F}}_1, \overline{\mathbf{F}}_2, \overline{\mathbf{F}}_3)^T$ with $\mathbf{F}_i=(\overline{F}^{i}_{1},\cdots,\overline{F}^{1}_{N_{\Lambda_{i}}})$,
%then the system (\ref{galproj1}) can be compactly written as
%\begin{equation}\label{agp}
%\dot{\overline{\mathbf{a}}}=\overline{\mathbf{F}}(\overline{\mathbf{a}},t),
%\end{equation}
%with appropriate initial condition.

%%\paragraph{Intrusive PWR }
%In intrusive PWR, after performing the approximate Galekin projection explicitly,
%the system (\ref{agp}) can be decomposed as
%\begin{equation}\label{apsdecomp}
%\dot{\overline{a}}^i_{j}=\overline{F}_{j}^i(\overline{\mathbf{a}}_i,\overline{\mathbf{d}}_i,t),
%%\dot{\overline{\mathbf{a}}_i}=\overline{\mathbf{F}}_i(\overline{\mathbf{a}}_i,
%%\overline{\mathbf{d}}_i,t),
%\end{equation}
%where $\overline{\mathbf{d}}_1 = P^{2,1}(\overline{\mathbf{a}}_2)$,
%$\overline{\mathbf{d}}_2 = (\overline{\mathbf{a}}_1, \overline{\mathbf{a}}_3)$
%and $\overline{\mathbf{d}}_3=P^{2,1}(\overline{\mathbf{a}}_2)$ are the decoupling
%vectors (here we overloaded notation for $P^{2,i}(\overline{\mathbf{a}}_2)$ to
%imply the coefficients in expansion (\ref{proj})). Note that the decomposition of
%system (\ref{agp}) is naturally induced by the decomposition of the original
%system (\ref{complexsys}). Fixed step waveform relaxation can then be applied to
%solve the decomposed system iteratively.

\paragraph{Non-Intrusive PWR }
The basic idea is to apply PCM at subsystem level at each PWR iteration, use gPC to represent the probabilistic waveforms and iterate among subsystems using these waveforms. Recall that in standard PCM approach, the coefficients $\overline{a}_{m}^i(t)$ are obtained by calculating the
integral
\begin{eqnarray}
\overline{a}_{m}^i(t)&=&\int x_{i}^{\Lambda_i}(t,\mathbf{\xi})\Psi_m^{\Lambda_i}(\mathbf{\xi})\mathbf{w}(\mathbf{\xi})d\mathbf{\xi}\label{z4approx}
\end{eqnarray}
The integral above is typically calculated by using a quadrature formula and
repeatedly solving the $i^{th}$ subsystem over an appropriate collocation grid %$\mathcal{C}^i$
%\begin{equation}\label{fullgirdgen}
$\mathcal{C}^i(\Sigma_i)=\mathcal{C}^i(\Lambda_i)\times\mathcal{C}^i(\Lambda_i^c)$,
%\end{equation}
where, $\mathcal{C}^i(\Lambda_i)$
is the collocation grid corresponding to parameters $\Lambda_i$ (and let $l_s$
be the number of grid points for each random parameter in $\Lambda_i$),
$\mathcal{C}^i(\Lambda_i^c)$ is the collocation grid corresponding to
parameters $\Lambda^c_i$ (and let $l_c$ be the number of grid points for each
random parameter in  $\Lambda^c_i$ ). Since, the behavior of $i^{th}$ subsystem is
weakly affected by the parameters $\Lambda^c_i$, we can take a sparser grid
in $\Lambda^c_i$ dimension, i.e.
%\begin{equation}\label{maxcond}
$l_c<l_s$.
%\end{equation}

Here we outline key steps of an algorithm for fixed-step non intrusive PWR.
\begin{itemize}
\item Step 1: (Initialization of the relaxation process with no coupling effect): Set $I=1$, guess an initial waveform $x_i^0(t)$ consistent with initial condition. Set
$\mathbf{d}_{1}^1=x^0_{2},\quad \mathbf{d}_{2}^1=(x^0_{1},x^0_3),\quad \mathbf{d}_{3}^1=x^0_2,$ and solve
\begin{equation}
  \dot{x}_i^1=f_i(x_i^1,\mathbf{d}_{i}^1(t),\xi_i,t),
\end{equation}
with an  initial condition $x^1_i(0)=x_i^0(0)$ on a collocation grid $\mathcal{C}^i(\Lambda_i)$. Determine the gPC expansion $x_{i}^{\Lambda_i,1}(t,\cdot)$ by computing the expansion coefficients from the quadrature formula (\ref{z4approx}).
\item Step 2: (Initialization of the relaxation process, incorporating first level of coupling effect): Set $I =2$ and $\mathbf{d}_{1}^2=x_{2}^{\Lambda_2,1},\quad \mathbf{d}_{2}^2=(x_{1}^{\Lambda_1,1},x_{3}^{\Lambda_3,1}),\quad \mathbf{d}_{3}^2=x_{2}^{\Lambda_2,1}$
and solve
\begin{equation}
  \dot{x}_i^2=f_i(x_i^2,\mathbf{d}_{i}^2(t,\cdot),\xi_i,t),
\end{equation}
over a collocation grid $\mathcal{C}^i(\Sigma_i)$ to obtain
$x_i^{\Sigma_i,2}(t,\cdot)$. From now on we shall denote the solution
of the $i^{th}$ subsystem at $I^{th}$ iteration by  $x_{i}^{\Sigma_i,I}$.

\item Step 3 (Analyzing the decomposed system at the I-th iteration):
Set $\mathbf{d}_{1}^I=P^{2,1}(x_{2}^{\Sigma_2,I-1}),\quad \mathbf{d}_{2}^I=(x_{1}^{\Sigma_1,I-1},x_{3}^{\Sigma_3,I-1}),\mathbf{d}_{3}^I=P^{2,3}(x_{2}^{\Sigma_2,I-1})$ and solve
\begin{equation}
  \dot{x}_i^I=f_i(x_i^I,\mathbf{d}_{i}^I(t,\cdot),\xi_i,t),
\end{equation}
over a collocation grid $\mathcal{C}^i(\Sigma_i)$ and obtain the expansion $x_i^{\Sigma_i,I}(t,\cdot)$.

\item Step 4 (Iteration) Set $I = I + 1$ and go to step 5 until satisfactory convergence has been achieved.
\end{itemize}

\paragraph{Convergence of PWR} Convergence of the intrusive and non-intrusive PWR
algorithm is guaranteed under very mild assumptions on the system (\ref{complexsys}).
As in the case of deterministic waveform relaxation \cite{waveform1}, the PWR converges
if the system (\ref{complexsys}) is Lipschitz. Details of the proof can be
found in \cite{pwr_cdc10}.

\begin{figure}[th!]
\begin{center}
\includegraphics[width=0.45\hsize]{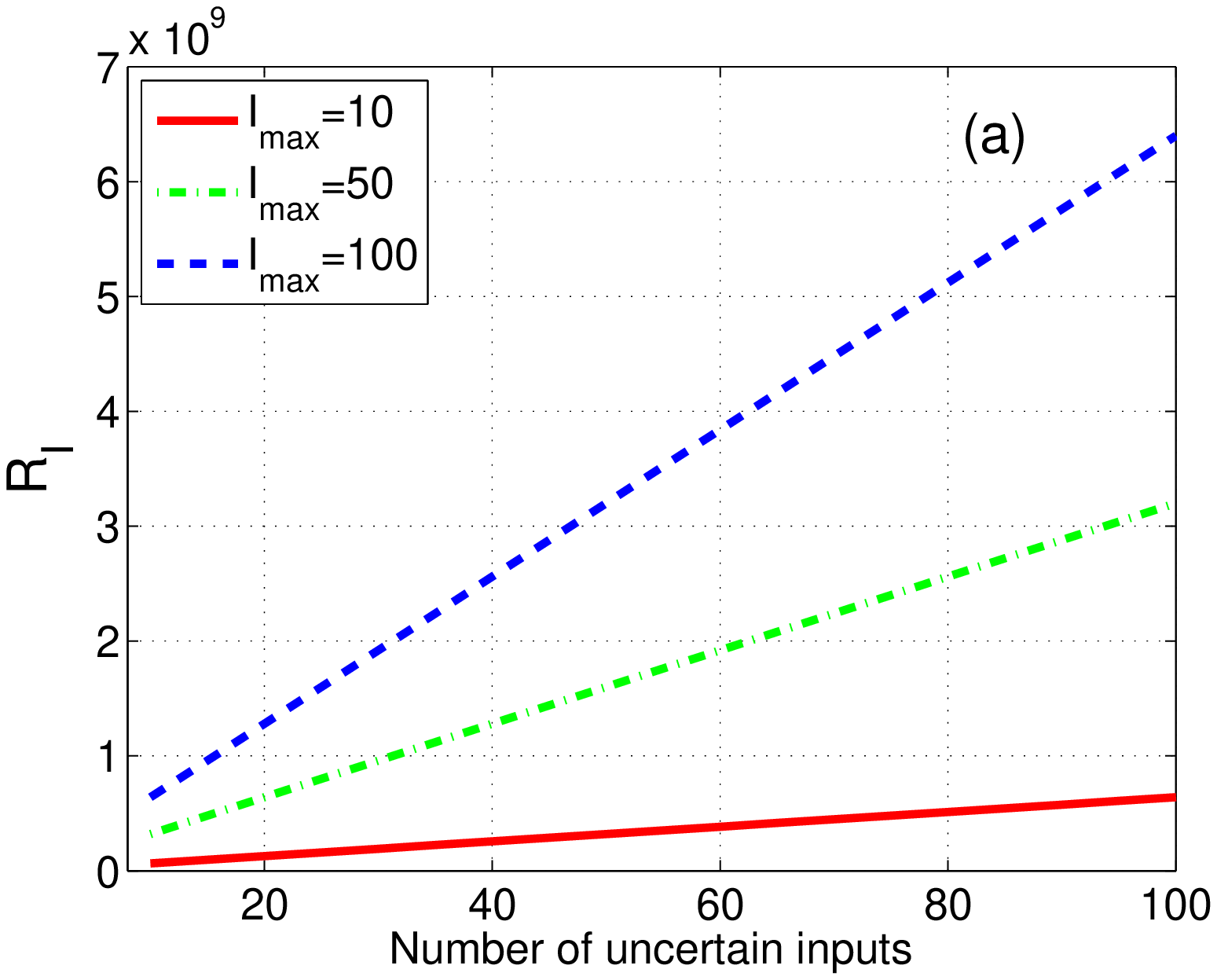}\hfill
\includegraphics[width=0.45\hsize]{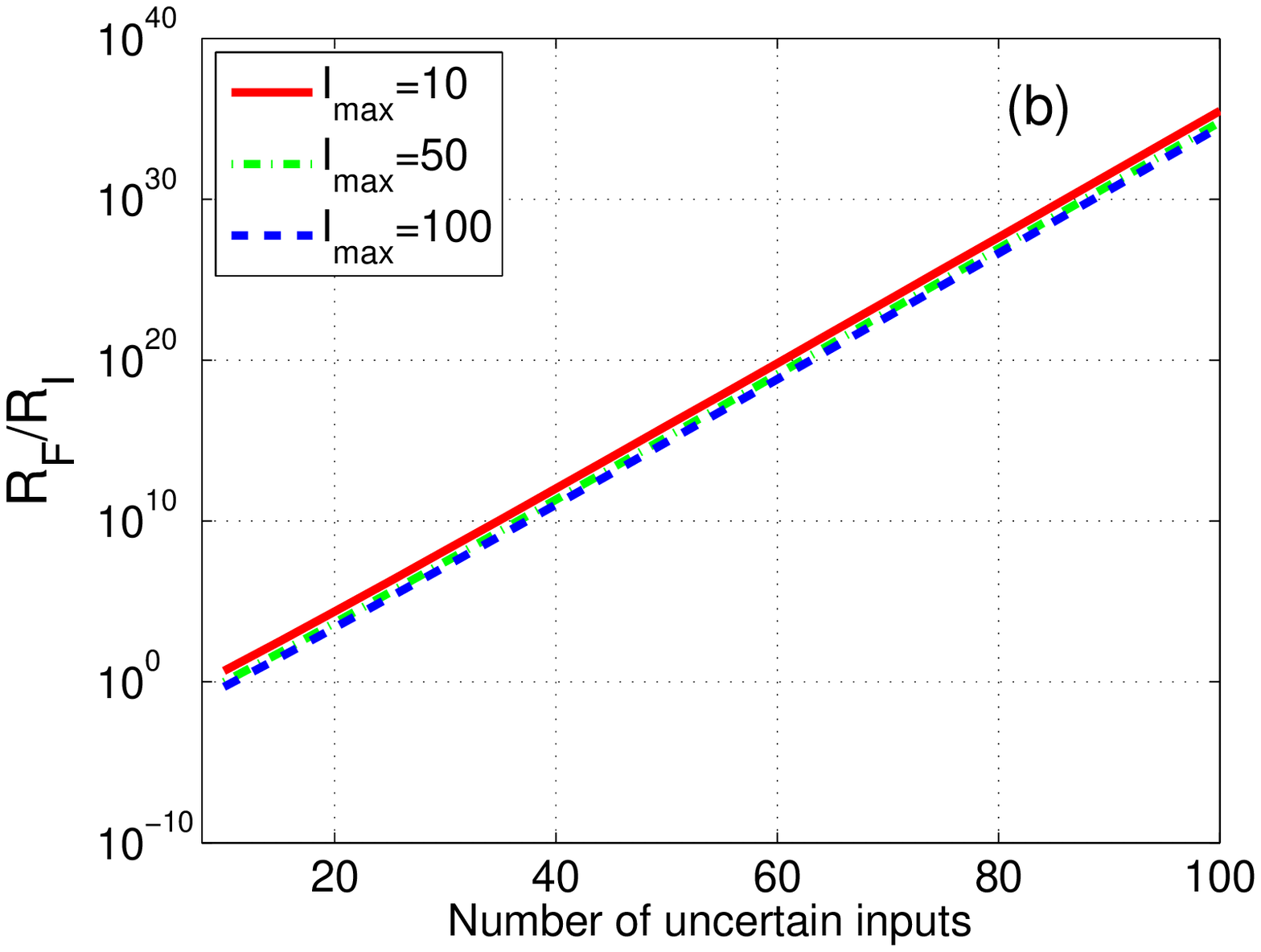}
\caption{(a) Projected scaling of PWR algorithm is linear with respect to
the number of uncertain inputs when system can be decomposed in an array of nearest neighbor coupled subsystems. $I_{\max}$ is the anticipated number of waveform relaxation iterations. (b) Projected computational cost ratio of full grid collocation and PWR methods $\mathcal{R}_F/\mathcal{R}_I$ plotted versus number of uncertain inputs for the case when clusters are all-to-all coupled. PWR outperforms full grid collocation method by several orders of magnitude.}\label{fig:scaling_projection}
\end{center}
\end{figure}

\paragraph{Scalability of PWR} The scalability of PWR relative to full grid PCM is shown in Figure \ref{fig:scaling_projection}, where the ratio $\mathcal{R}_F/\mathcal{R}_I$ indicates the computation gain over standard full grid approach applied to the system as a whole. Here $\mathcal{R}_F=l^{p}$ is the number of deterministic runs of the complete system (\ref{complexsys}), comprises of $m$ subsystems each with $p_i,i=1,\cdots,m$ uncertain parameters, such that $p=\sum_{i=1}^mp_i$ and $l$ denotes the level of full grid. Similarly, $\mathcal{R}_I=1+\sum_{i=1}^ml_s^{p_i}+I_{\mbox{max}}(\sum_{i=1}^m l_s^{p_i}\bigotimes_{j\neq i}l_c^{p_j})$ is the total computational effort with PWR algorithm, where  $I_{\max}$ is the number of PWR iterations. Clearly, the advantage of PWR becomes evident as the number of subsystems $m$ and parameters in the network increases.

\subsection{Numerical Example}

\subsubsection{Coupled Oscillators}
\begin{figure}[th!]
\begin{center}
  \begin{tabular}{cc}
  \includegraphics[width=0.45\hsize]{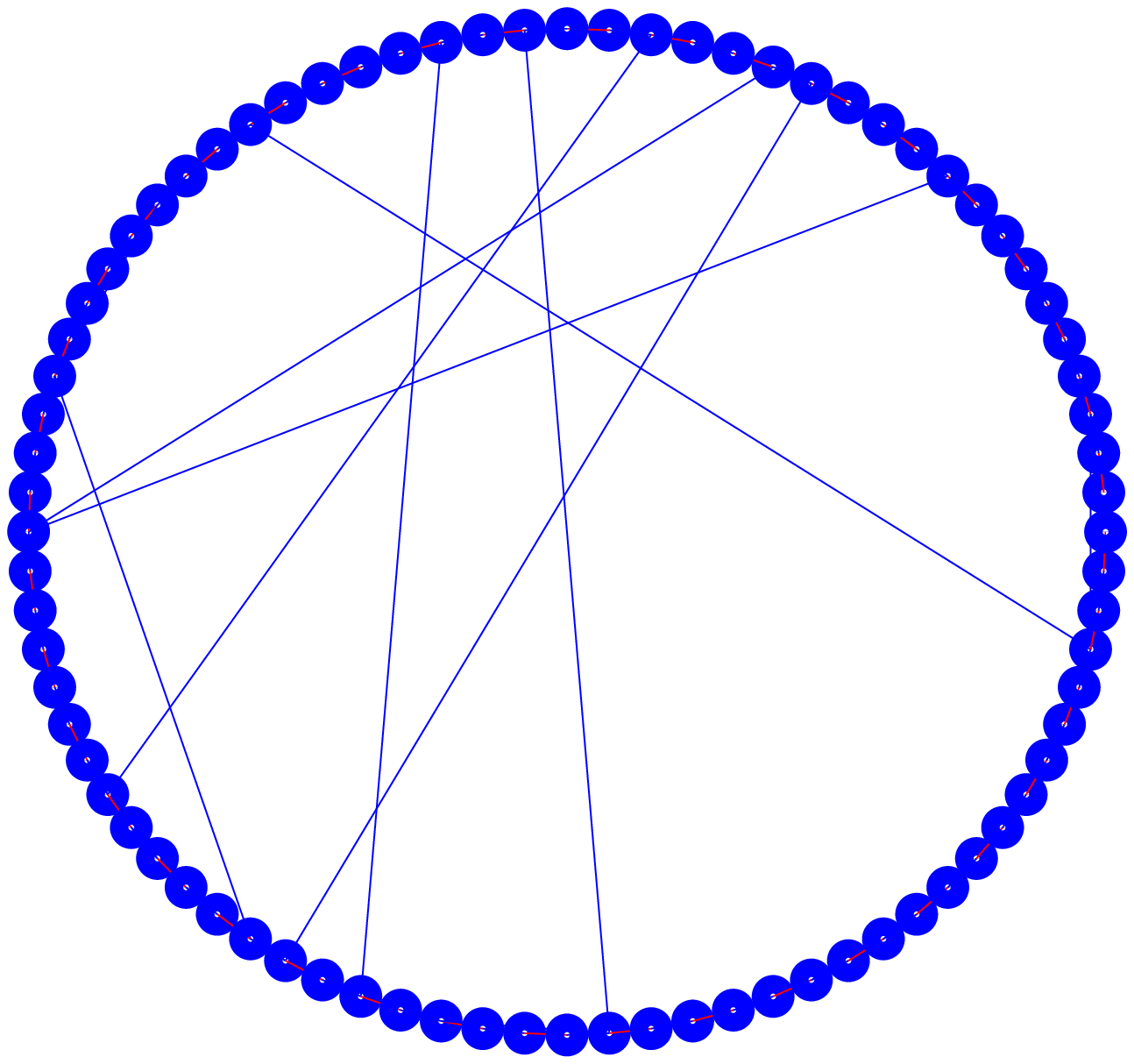}  &
  \includegraphics[width=0.5\hsize]{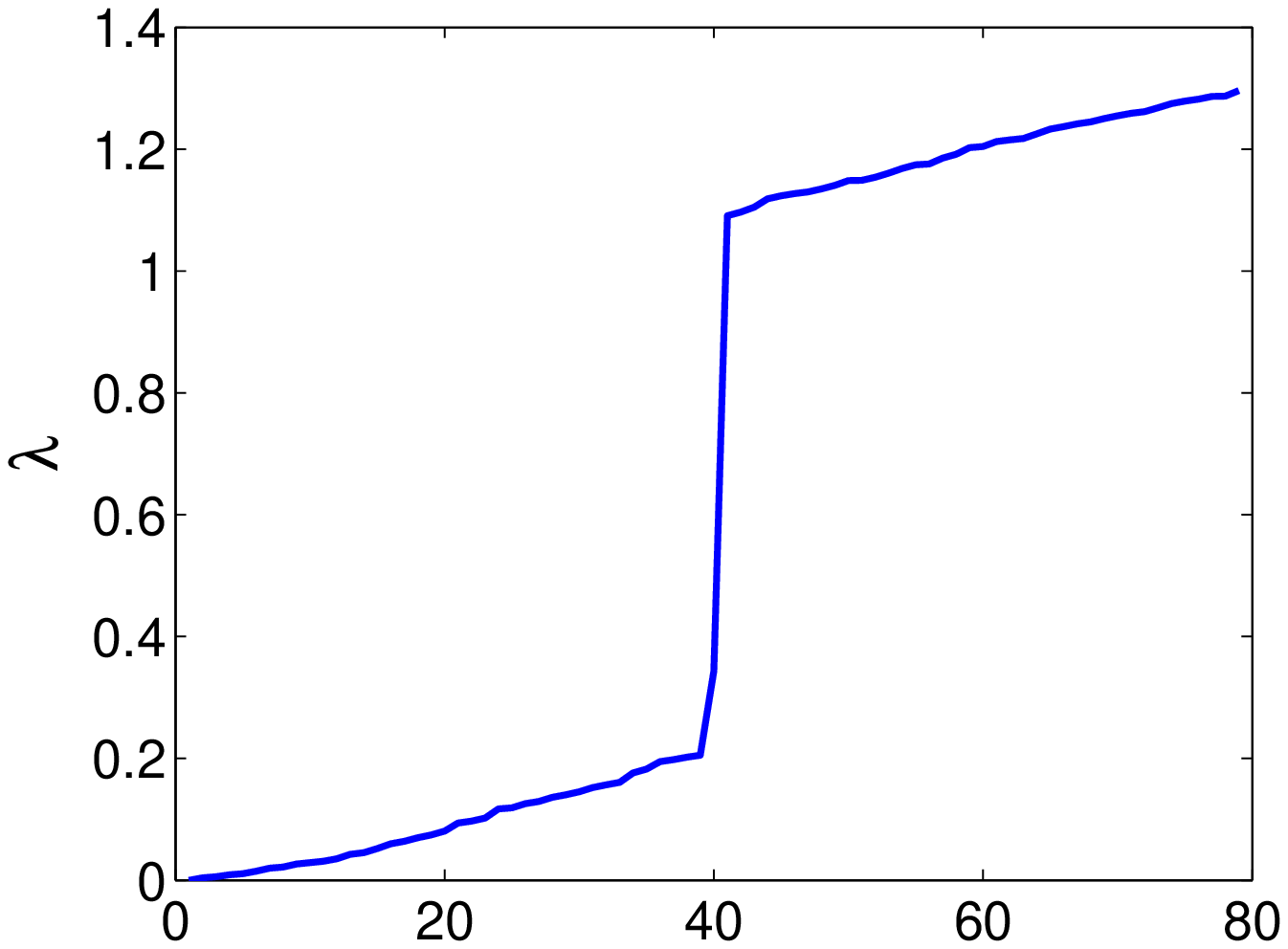}\\
  \footnotesize{(a)} & \footnotesize{(b)}
    \end{tabular}
\caption{(a) shows a network  of $N=80$ phase only oscillators. (b) shows spectral gap in eigenvalues of normalized graph Laplacian, that reveals that there are $40$ weakly interacting subsystems.}\label{fig1}
\end{center}
\end{figure}

In this section we consider a coupled phase only oscillator system which is governed by
\begin{equation}\label{ocslliator}
\dot{x}_i=\omega_i+\sum_{j=1}^N K_{ij} \sin(x_j-x_i),\qquad i=1,\cdots,N,
\end{equation}
where, $N=80$ is the number of oscillators, $\omega_i,i=1,\cdots,N$ is the angular frequency of oscillators and $K=[K_{ij}]$ is the coupling matrix. The frequencies $\omega_i$ of every alternative oscillator i.e. $i=1,3,\cdots,79$ is assumed to be uncertain with a Gaussian distribution with $20\%$ tolerance (i.e. with a total $p=40$ uncertain parameters); all the other parameters are assumed to take a fixed value. We are interested in the distribution of the synchronization parameters, $R(t)$ and phase $\phi(t)$, defined by $R(t)e^{\phi(t)}=\frac{1}{N}\sum_{j=1}^N e^{ix_j(t)}$. Figure \ref{fig1} shows the topology of the network of oscillators (left panel), along with the eigenvalue spectrum of the graph Laplacian (right panel). The spectral gap at $40$, implies $40$ weakly interacting subsystems in the network. We use spectral clustering algorithm given in \cite{specgraph} to identify these subsystems.

\begin{figure}[t!]
\begin{center}
\includegraphics[width=0.8\hsize]{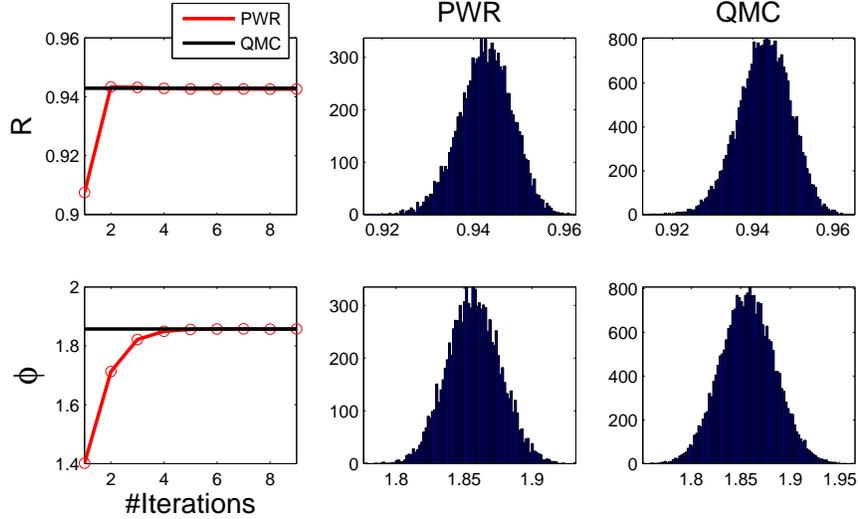}\\
\caption{Convergence of mean of the magnitude $R(t)$ and phase $\phi(t)$, and the respective histograms at $t=0.5$. Also shown on right are the histograms obtained from PWR and QMC approaches.}\label{fig2}
\end{center}
\end{figure}

Figure \ref{fig2} shows UQ results obtained by application of PWR to the decomposed system with $l_s=5$, $l_c=2$ and $P=5$. We make comparison with QMC, in which the complete system (\ref{ocslliator})  is solved at $25,000$  Sobol points \cite{Kuo1}. Remarkably the PWR converges in $4-5$ iterations giving very similar results to that of QMC. It would be infeasible to use full grid collocation for the network as a  whole, since even with lowest level of collocation grid, i.e. $l=2$ for each parameter, the number of samples required become $N=2^{40}=1.0995e+012!$.

\subsubsection{Building Example}
We revisit the building thermal network model example studied in section \ref{sec:buildeg}, and  consider the problem of computing uncertainty in building
energy consumption due to parametric uncertainty (e.g., wall thermal conductivity and thermal
capacitance),  and  external (e.g., solar radiation) and internal
(occupant load) time varying stochastic disturbances.  To account for time varying uncertain processes in above framework, we will employ
Karhunen Loeve (KL) expansion \cite{ghanem1991}. The KL expansion allows representation
of second order stochastic processes as a sum of random variables. Hence, both parametric and time varying uncertainties can be treated in terms of random variables.

\begin{figure}[t!]
\begin{center}
\includegraphics[width=0.7\hsize]{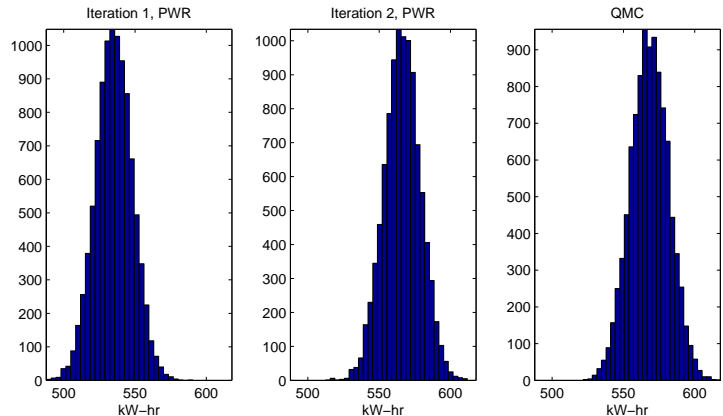}
\end{center}
\caption{Histogram of building energy computation for two iterations in PWR.
Also shown is the corresponding histogram obtained by QMC for comparison.}
\label{Fig:buildPWR}
\end{figure}

To illustrate the PWR approach we consider the effect of $14$ random variables (comprising of wall thermal capacitance and conductivity and random variables obtained in a KL representation of internal and external load) in the thermal network model on the building energy consumption. Recall that the $68$ state thermal network model admits a decomposition into $23$ subsystems (see section \ref{sec:buildeg} for details). Figure \ref{Fig:buildPWR} shows  results of application of PWR on the decomposed network model. As is evident, the iterations converge rapidly in two steps with a distribution close to that obtained from QMC (using a 25000-sample Sobol sequence) applied to the thermal network model as a whole.

\subsection{Extensions}
Several questions need to be further investigated, these include: how the choice of parameters associated with PWR algorithm affects its rate of convergence and the approximation error. In order to exploit multiple time scales that may be present in a system, multigrid extension \cite{mg} of PWR will also be desirable.

\section{Sensor tasking for Uncertainty Reduction in Multitarget Search and Tracking Applications}\label{sec:algorithm}
In this section we address the problem of managing uncertainty and complexity of planning and executing Intelligence Surveillance Reconnaissance (ISR) missions using a network of sensor resources like Unmanned Arial Systems (UAS) or a network of cameras. In particular we consider the problem of designing dynamics of mobile sensors for multi-target search and tracking applications. In such applications, it is important to design uniform coverage dynamics such that there is little overlap of the sensor footprints and little space left between the sensor footprints. In other words, the sensor footprints must be uniformly distributed so that it becomes difficult for a target to evade detection. These requirements motivated the development of the Spectral Multiscale Coverage (SMC) algorithm described in \cite{smc_cdc09} and \cite{smc_physd09}. For the search of a stationary target, the uncertainty in the position of the target can be specified in terms of a fixed probability distribution. The SMC algorithm makes the sensors move so that points on the sensor trajectories uniformly sample this stationary probability distribution. Uniform coverage dynamics coupled with sensor observations helps to reduce the uncertainty in the position of the target. In \cite{MAS}, it has been demonstrated that in the presence of various uncertainties, uniform coverage based search strategies outperform lawnmower-type search strategies. An extension of the SMC algorithm - Dynamic Spectral Multiscale Coverage (DSMC) - has been developed in \cite{dsmc_cdc10} for the search of moving targets.

Some representative papers that deal with the problem of coverage/sampling are \cite{Howard02mobilesensor, bullo_coverage, leonard_lekien_coverage, hussein_coverage, ocean_sampling_leonard}. The term ``coverageï'' can mean slightly different things to different authors. For example, in \cite{Howard02mobilesensor, bullo_coverage, leonard_lekien_coverage}, ``coverage'' is more a static concept. i.e., it is a measure of how a static configuration of agents covers a domain or samples a probability distribution. In \cite{hussein_coverage} and \cite{ocean_sampling_leonard}, the term �coverage'' is more of a dynamic concept and is a measure of how well the points on the trajectories of the sensor trajectories cover a domain. That is, coverage gets better and better as every point in the domain is visited or is close to being visited by an agent. The notion of coverage we use is closer to that in \cite{hussein_coverage} and \cite{ocean_sampling_leonard}. Moreoever, we use the notion of ``uniform coverage'' which we quantify using metrics inspired by the ergodic theory of dynamical systems. The behavior of an algorithm that attempts to achieve uniform coverage is going to be inherently multiscale. By this we mean that, features of large size are guaranteed to be detected first, followed by features of smaller and smaller size.

\subsection{Uniform Coverage Algorithms}
A system is said to exhibit {\it ergodic dynamics} if it visits every subset of the phase space with a probability equal to the {\it measure} of that subset. For good coverage of a stationary target, this translates to requiring that the amount of time spent by the mobile sensors in an arbitrary set be proportional to the probability of finding the target in that set. For good coverage of a moving target, this translates to requiring that the amount of time spent in certain ``tube sets" be proportional to the probability of finding the target in the ``tube sets".  We assume a model for the motion of the targets to construct these ``tube sets" and define appropriate metrics for coverage. (The model for the target motion can be approximate and the dynamics of targets for which we don't have precise knowledge can be captured using stochastic models). Using these metrics for uniform coverage, we derive centralized feedback control laws for the motion of the mobile sensors.

\subsubsection{Coverage dynamics for stationary targets}

\begin{figure}[th!]
  \begin{center}
  \subfigure[Time,t=0.0]{\includegraphics[width=0.35\hsize]{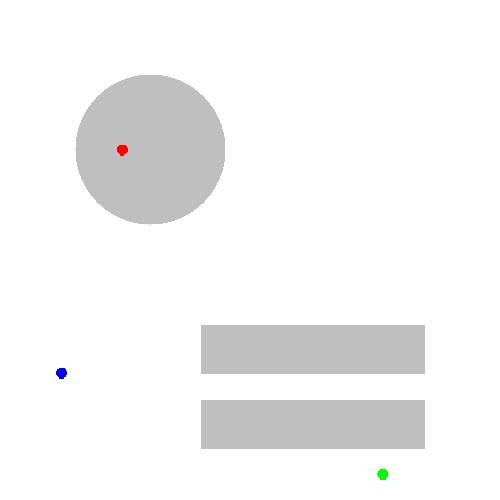}}\hspace*{1.5cm} \subfigure[Time,t=6]{\includegraphics[width=0.35\hsize]{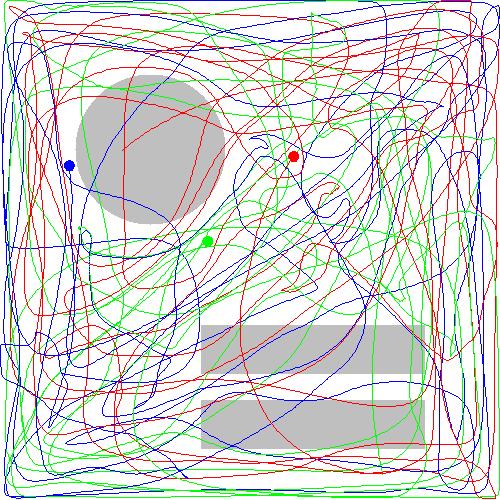}}\\
  \subfigure[Time,t=12]{\includegraphics[width=0.35\hsize]{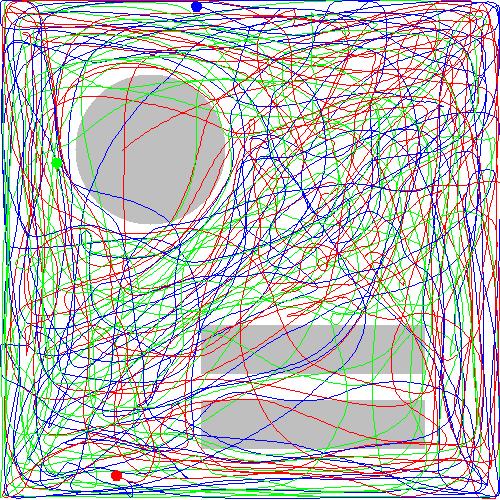}}\hspace*{1.5cm}
  \subfigure[Time,t=18]{\includegraphics[width=0.35\hsize]{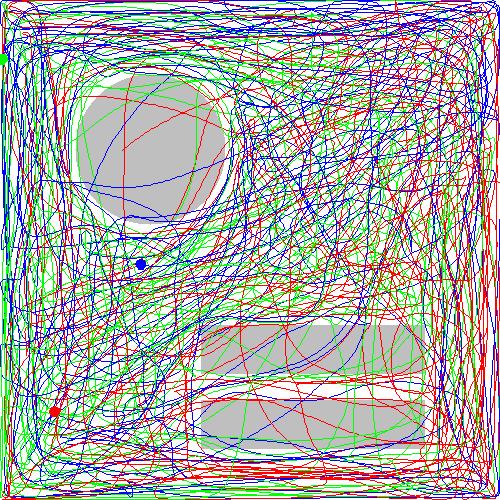}}
  \end{center}
  \caption{Snapshots at various times of the agent trajectories generated by the SMC algorithm to uniformly cover an irregular domain. One can observe the multiscale nature of the algorithm. The spacing between the trajectories becomes smaller and smaller as time proceeds.}
\label{Fig:SMC_1storder}
\end{figure}

There are N mobile agents and we assume that they move either by first-order or second-order dynamics. We need an appropriate metric to quantify how well the trajectories are sampling a given probability distribution $\mu$. We assume that $\mu$ is zero outside a rectangular domain $U \subset \mathbb{R}^n$ and that the agent trajectories are confined to the domain $U$. For a dynamical system to be ergodic, we know that the fraction of the time spent by a trajectory in a subset must be equal to the measure of the set. Let $B(x, r) = \{y : \|y- x \| \leq r \}$ be a spherical set and $\chi_{(x,r)}$ be the indicator function corresponding to the set $B(x, r)$. Given trajectories $x_j : [0, t] \rightarrow \mathbb{R}^n$, for $j = 1, 2, ..., N$, the fraction of the time spent by the agents in the set $B(x, r)$ is given as

\begin{equation}
d^t(x,r) = \frac{1}{Nt} \sum_{j=1}^{N} \int_{0}^{t} \chi_{(x,r)}\left ( x_j(\tau) \right ) d \tau.
\end{equation}
The measure of the set $B(x, r)$ is given as
\begin{equation}
\bar{\mu}(x,r) = \int_{U} \mu(y) \chi_{(x,r)}(y) dy.
\end{equation}
For ergodic dynamics, we must have
\begin{equation}
\lim_{t \rightarrow \infty} d^t(x,r) = \bar{\mu}(x,r)
\end{equation}
Since the equality above must be true for almost every point $x$ and all radii $r$, this motivates defining the metric
\begin{equation}
E^2(t) = \int_{0}^{R} \int_{U} \left( d^t(x,r) - \bar{\mu}(x,r) \right)^2 dx dr.
\end{equation}
$E(t)$ is a metric that quantifies how far the fraction of the time spent by the agents in spherical sets is from being equal to the measure of the spherical sets. Now consider the distribution $C^t$ defined as
\begin{equation}
C^t(x) = \frac{1}{Nt} \sum_{j=1}^{N} \int_{0}^{t} \delta \left( x - x_j(\tau) \right) d \tau.
\end{equation}
Let $\phi(t)$ be the distance between $C^t$ and $\mu$ as given by the Sobolev space norm of negative index $H^{-s} \text{ for } s= \frac{n+1}{2}$. i.e.,
\begin{equation}
\begin{split}
\phi^2(t) &= \| C^t - \mu \|^2_{H^{-s}} = \sum_{K} \Lambda_k | s_k(t) |^2, \\
\text{where } s_k(t) &= c_k(t) - \mu_k, \quad \Lambda_k = \frac{1}{\left( 1 + \|k \|^2 \right)^s}, \\
c_k(t) &= \langle C^t, f_k \rangle, \quad \mu_k = \langle \mu, f_k \rangle.
\end{split}
\end{equation}
Here, $f_k$ are Fourier basis functions with wave-number vector $k$. The metric $\phi^2(t)$ quantifies how much the time averages of the Fourier basis functions deviate from their spatial averages, but with more importance given to large-scale modes than the small-scale modes. It can be shown that the two metrics $E(t)$and $\phi(t)$ are equivalent. Now consider the case where the sensors are moving by first-order dynamics described by
\begin{equation}
\dot{x}_j(t) = u_j(t).
\end{equation}
The objective is do design feedback laws $u_j(t) = F_j(x)$ so that the agents have ergodic dynamics. We formulate a model predictive control problem where we try to maximize the rate of decay of the coverage metric $\phi^2(t)$ at the end of a short time horizon and we derive the feedback law in the limit as the size of the receding horizon goes to zero. We take the cost-function to be the first time-derivative of $\phi^2(\tau)$ at the end of the horizon $\left[ t, t + \Delta t \right]$. i.e.,
\begin{equation}\label{eq:1stordercost}
C(t, \Delta t)  =  \sum_{K} \Lambda_k s_k(t+ \Delta t) \dot{s}_k(t + \Delta t).
\end{equation}
The feedback law in the limit as the receding horizon $\Delta t$ goes to zero, is given as
\begin{equation}
\begin{split}
u_j(t) & = -u_{max} \frac{B_j}{ \| B_j(t) \|_2}, \\
\end{split}
\label{smc_1storder:fdbk}
\end{equation}
where, $B_j(t) = \sum_{k} \Lambda_k s_k(t) \nabla f_k(x_j(t))$ and $\nabla f_k(.)$ is the gradient vector field of the Fourier basis functions $f_k$. Figure \ref{Fig:SMC_1storder} shows snapshots of trajectories generated by the SMC algorithm to uniformly cover an irregular domain. From Figure \ref{Fig:SMC_1storder}, one can observe the multiscale nature of the algorithm. The spacing between the trajectories becomes smaller and smaller as time proceeds.

%\begin{figure}
%\begin{center}
%\begin{tabular}{c c }
%\subfigure[Time,t=0.0]{\includegraphics[scale=0.18]{SMC_1storder_T00000.jpg}}&
%\subfigure[Time,t=6.0]{\includegraphics[scale=0.18]{SMC_1storder_T06000.jpg}}\\
%\subfigure[Time,t=12.0]{\includegraphics[scale=0.18]{SMC_1storder_T12000.jpg}}&
%\subfigure[Time,t=18.0]{\includegraphics[scale=0.18]{SMC_1storder_T18000.jpg}}\\
%\end{tabular}
%\end{center}
%\caption{Snapshots at various times of the agent trajectories generated by the SMC algorithm to uniformly cover an irregular domain. One can observe the multiscale nature of the algorithm. The spacing between the trajectories becomes smaller and smaller as time proceeds.}
%\label{Fig:SMC_1storder}
%\end{figure}

\subsubsection{Coverage dynamics for moving targets}
Let the target motion model be described by a deterministic set of ODE's
\begin{equation}
\dot{z}(t) = v(z(t), t),
\end{equation}
where, $z(t)\in U$, $U\subset \mathbb{R}^2$ being the region in which the target motion is confined to over a period $[0,T_f]$ of interest. Let $T$ be the corresponding mapping that describes the evolution of the target position. i.e., if the target is at point $z(t_0)$ at time $t = t_0$, its position at time $t = t_f$ is given by $z(t_f) = T( z(t_0) , t_0, t_f)$. Given a set $A\subset U$ its inverse image under the transformation $T(.,t_0, t_f)$ is given as
\begin{equation}
T^{-1}(., t_0, t_f)(A) = \{ y: T(y, t_0, t_f) \in A \}.
\end{equation}
The initial uncertainty in the position of the target is specified by the probability distribution $\mu(0, x) = \mu_0(x)$. Let $[P^{t_0, t_f}]$ be the family of Perron-Frobenius operators corresponding to the transformations $T(.,t_0, t_f)$. i.e.,
\begin{equation}
\int\limits_{A} [P^{t_0,t_f}] \mu(t_0, y) dy = \int\limits_{A} \mu(t_f, y) dy = \int \limits_{T^{-1}(.,t_0, t_f)(A)} \mu(t_0, y) dy.
\end{equation}
At time $t$, consider the spherical set $B(x,r)$ with radius $r$ and center at $x$. Now consider the corresponding tube set given as
\begin{equation}\label{eq:tubeset}
H^t(B(x,r)) = \left \{ (y, \tau) : \tau \in [0,t] \text{ and } T(y, \tau, t) \in B(x,r) \right \}.
\end{equation}
%\begin{figure}
%\begin{center}
%\subfigure{\includegraphics[scale=0.35]{cylinder_set.jpg}}
%\end{center}
%\caption{The tube set $H^t(B(x,r))$ is a subset of the extended space-time domain and is the union of the sets $ T^{-1}(. , \tau, t)(B(x,r)) \times \{\tau\}$ for all $\tau \in [0,t]$. Note that there is no particular reason that the target trajectories going back in time should diverge. They could just as well converge and then the set of initial target conditions leading to $B(x,r)$ could get smaller and smaller. This still does not change the observation in \eqref{consmass}. }
%\label{Fig:tube_set}
%\end{figure} (See Figure \ref{Fig:tube_set})
The tube set $H^t(B(x,r))$  is a subset of the extended space-time domain and is the union of the sets $T^{-1}(., \tau, t)(B(x,r)) \times \{\tau \}$ for all $\tau \in [0,t]$. This tube set can be thought of as the set of all points in the extended space-time domain that end up in the spherical set $B(x,r)$ at time $t$ when evolved forward in time according to the target dynamics. Note that the probability of finding a target within any time slice of the tube set is the same. i.e.,
\begin{equation} %\label{consmass}
\begin{split}
\mu(\tau_1, T^{-1}(. , \tau_1, t)(B(x,r))) &= \mu(\tau_2, T^{-1}(. , \tau_2, t)(B(x,r)))  \\
&= \mu(t, B(x, r)),\notag
\end{split}
\end{equation}
$\text{for all } \tau_1, \tau_2 \leq t$. This is because none of the possible target trajectories either leave or enter the tube set $H^t(B(x,r))$. Let the sensor trajectories be $x_j: [0, t] \rightarrow \mathbb{R}^2$ for $j=1,2,..,N$. The fraction of the time spent by the sensor trajectories $(x_j(t), t)$ in the tube set $H^t(B(x,r))$ is given as
\begin{equation}
\begin{split}
d^t(x,r) &= \frac{1}{N t} \sum_{j=1}^{N} \int_{0}^{t} \chi_{T^{-1}(., \tau, t)(B(x,r))} \left(x_j(\tau) \right) d \tau \\
& = \frac{1}{N t} \sum_{j=1}^{N} \int_{0}^{t} \chi_{B(x, r)} \left( T( x_j(\tau), \tau, t) \right)  d \tau.
\end{split}
\label{frac:tube}
\end{equation}
$\chi_{B(x,r)}$ is the indicator function on the set $B(x,r)$.
It might appear that the fraction $d^t(x,r)$ is hard to compute. But it turns out that this fraction can be computed as the spherical integral
\begin{equation}
d^t(x, r) = \int \limits_{B(x,r)} C^t(y) dy,
\label{frac:sphint}
\end{equation}
of a distribution,
\begin{equation}
C^t(x) = \frac{1}{N t} \sum_{j=1}^{N} \int_{0}^{t} P^{\tau, t} \delta_{x_j(\tau)}(x) d\tau,
\label{equ:coveragedistribution}
\end{equation}
which we refer to as the {\it coverage distribution}. Here $\delta_{x_j(\tau)}$ is the delta distribution with mass at the point $x_j(\tau)$. The {\it coverage distribution} $C^t$ can be thought of as the distribution of points visited by the mobile sensors when evolved forward in time according to the target dynamics. In \cite{dsmc_cdc10} we describe a iterative procedure to numerically compute an approximation to the coverage distribution $C^t$.

\begin{figure}[t!]
  \begin{center}
    \subfigure[Time, t=0.0]{\includegraphics[width=0.22\hsize]{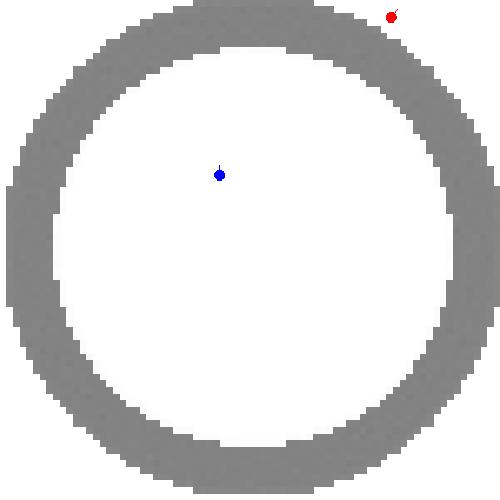}}\hfill \subfigure[ Time, t=0.6]{\includegraphics[width=0.25\hsize]{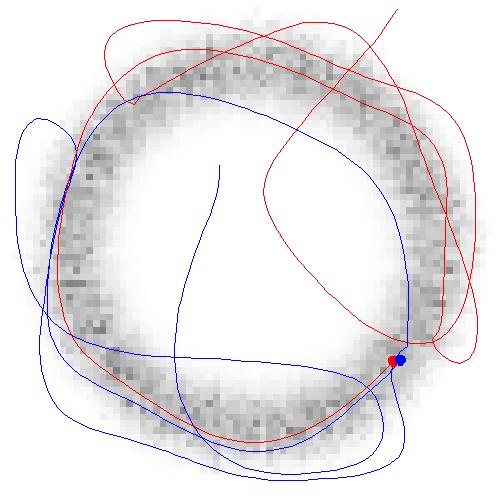}}\hfill
    \subfigure[Time, t=1.2]{\includegraphics[width=0.25\hsize]{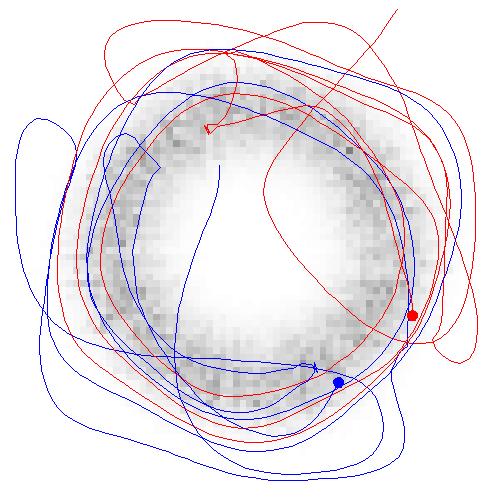}}\hfill
    \subfigure[Time, t=1.2]{\includegraphics[width=0.25\hsize]{DSMC_1storder_T00600.jpg}}\hspace*{1cm}
    \subfigure[Time, t=1.8]{\includegraphics[width=0.25\hsize]{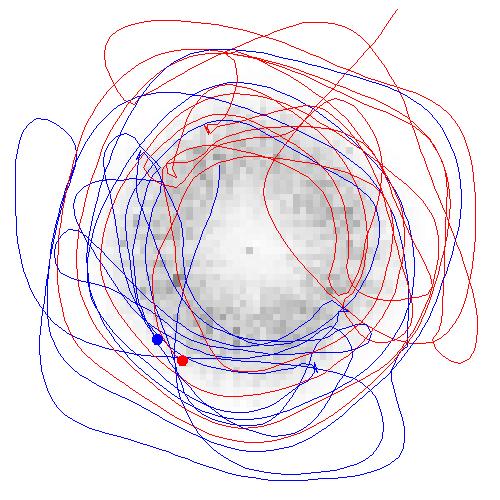}}\hspace*{1cm}
  \end{center}
  \caption{Snapshots at various times of the sensor trajectories generated by the DSMC algorithm to search for a moving target. The dynamics of the target is such that it moves toward a central point with a constant speed. The initial prior probability distribution is a uniform distribution on a ring of finite width. The grey blob in the pictures above represent the evolving uncertainty in the position of the moving target.}
\end{figure}

%\begin{figure}
%\begin{center}
%\begin{tabular}{c c }
%\subfigure[Time,t=0.0]{\includegraphics[scale=0.18]{DSMC_1storder_T00000.jpg}}&
%\subfigure[Time,t=0.6]{\includegraphics[scale=0.18]{DSMC_1storder_T00300.jpg}}\\
%\subfigure[Time,t=1.2]{\includegraphics[scale=0.18]{DSMC_1storder_T00600.jpg}}&
%\subfigure[Time,t=1.8]{\includegraphics[scale=0.18]{DSMC_1storder_T00899.jpg}}\\
%\end{tabular}
%\end{center}
%\caption{Snapshots at various times of the sensor trajectories generated by the DSMC algorithm to search for a moving target. The dynamics of the target is such that it moves toward a central point with a constant speed. The initial prior probability distribution is a uniform distribution on a ring of finite width. The grey blob in the pictures above represent the evolving uncertainty in the position of the moving target.}
%\label{Fig:DSMC_1storder}
%\end{figure}

For uniform sampling of the target trajectories, it is desirable that the fraction of the time spent by the sensor trajectories in the tube set must be close to the probability of finding a target trajectory in the tube which is given as
\begin{equation}
\mu(t, B(x,r)) = \int \limits_{B(x,r)} \mu(t, y) dy = \int \limits_{T^{-1}(., 0, t)(B(x,r))} \mu_0(y) dy.
\end{equation}
This motivates defining the metric
\begin{equation}
\begin{split}
\psi^2(t) = \| C^t - \mu(t, .) \|^2_{H^{-s}}
\end{split}
\end{equation}
Using the same receding horizon approach as described before for stationary targets, we get feedback laws similar to that in \eqref{smc_1storder:fdbk}. In an example simulation shown in  Figure \ref{Fig:DSMC_1storder}, we assume that the target motion is such that it moves towards a central point with a known fixed speed. The initial uncertainty in the target position is uniformly distributed in a ring of finite size centered around the origin.

\subsection{Extensions}
We have extended the work described here to the case when the target dynamics is stochastic. We have also extended the DSMC framework to deal with higher order target dynamics, to account for heterogeneity in their dynamics and to incorporate target prioritization in the coverage metric. Modifications of the algorithm to achieve decentralization in the case when there is limited communication between agents is being pursued. We are also extending the DSMC algorithm for sensors/vehicles with realistic dynamics (e.g. with under actuation) in an in an obstacle rich environment. In next section we consider this more difficult problem of optimizing motion of vehicle under different constraints.

\section{Robust Path Planning}\label{sec:Robust Path Planning}
We consider in this section the problem of trajectory planning for a vehicle with dynamics described by a realistic motion model and subject to multiple constraints  resulting from vehicle underactuation, actuator bounds, and environment obstacles. The performance of our trajectory-planning algorithm is first characterized through simulation followed by a brief discussion of recent experiments where the algorithm was used for real-life trajectory planning with a maxi-joker UAV.

\subsection{Problem Formulation}
Consider a vehicle with state $x\in X$ controlled using actuator
inputs $u\in U$, where $X$ is the state space and $U$ denotes the set
of controls. The vehicles dynamics is described by the function
$f:X\times U\rightarrow X$ defined by
\begin{align}\label{eq:f}
  \dot x = f(x, u) \text{\quad : vehicle dynamics model,}
\end{align}
which is used to evolve the vehicle state forward in time. In
addition, the vehicle is subject to constraints arising from actuator
bounds and obstacles in the environment. These constraints are expressed
through the $m$ inequalities
\begin{align}\label{eq:F}
  F_i(x(t))\ge 0 \text{\quad : constraints,}
\end{align}
for $i=1,...,m$.

The goal is to compute the optimal controls $u^*$ and final time $T^*$
driving the system from its initial state $x(0)$ to a given goal
region $X_g \subset X$, i.e.
\begin{align}\label{eq:J}
  \begin{split}
    & (u^*, T^*) = \arg\min_{u,T} \int_0^T C(x(t), u(t)) \rm{d} t, \\
    & \text{ subject to }\  \dot{x}(t) = f(x(t), u(t)), \\
    & \hspace{54pt} F_i(x(t)) \geq 0, \ i=1,...,m. \\
    & \hspace{54pt} x(T) \in X_g
  \end{split}
\end{align}
where $C:X \times U\rightarrow\mathbb{R}$ is a given cost
function. A typical cost function includes a time component and a
control effort component, i.e. $C(x, u)=1 + \lambda\|u\|^2$ where
$\lambda\ge 0$ is appropriate weighing factor.

\paragraph{Obstacle Constraints}
The vehicle operates in a workspace  that contains a
number of \emph{obstacles} denoted by
$\mathcal{O}_1, ...,\mathcal{O}_{n_o}$ with which the
vehicle must not collide. Typically, the vehicle state can be defined as $(q,v)$
consisting of its configuration $q\in \mathcal C$
and velocity $v \in \mathbb{R}^{n_v}$ ~\cite{La1991}. Assume that the vehicle
is occupying a region $\mathcal{A}(q,v)$, and let \textbf{prox}$({\mathcal A}_1,
\mathcal A_2)$ be the Euclidean distance between two sets
$\mathcal A_{1,2}$ that is negative in case of intersection.
Obstacle avoidance constraint in~\eqref{eq:F} can be written as
\begin{align}\label{eq:prox}
  F_1((q(t),v(t))) = \min_i  \textbf{prox}({\mathcal A}(q(t)), \mathcal
  O_i), \text{ for all } t\in[0,T].
\end{align}

\paragraph{Sensing}
We assume that the vehicle is equipped with a sensor measuring the
relative positions of obstacles, producing a set of points lying on the surface of
obstacles. In this work, we simplify the perception problem and
assume an on-board simultaneous localization and mapping (SLAM) algorithm for map updates.
The path planning algorithms developed in this section are then based only on the expected value of this map.

\subsection{Approach}
The problem~\eqref{eq:J} has no closed-form solution since both the
dynamics~\eqref{eq:f} and constraints~\eqref{eq:F} are nonlinear.
Gradient-based optimization is not suitable unless a good starting
guess is chosen since the constraints~\eqref{eq:F} impose many local
minima. In addition, special differentiation~\cite{ClLeStWo1998} is
required to guarantee convergence due to the non-smooth nature of the
constraints. An alternative is to discretize the vehicle state space
$X$ and transform the problem into a discrete graph search.
Such an approach suffers from the \emph{curse of dimensionality} due
to the exponential size of the search space (state dimension $\times$
trajectory epochs) and is limited to very simple problems.

In this paper we also employ a graph-based search but unlike
in standard discrete search, the vertices of the graph are \emph{sampled}
from the original continuous space $X$ and the edges correspond to
feasible trajectories, i.e. that satisfy the given
dynamics~\eqref{eq:f} and general constraints~\eqref{eq:F}.
Our approach is based on a recent methodology under active development
in the robotics community known as \emph{sampling-based motion
  planning} which includes the
\emph{rapidly-exploring random tree} (RRT)~\cite{La2006} and the
\emph{probabilistic roadmap} (PRM)~\cite{ChLyHuKaBuKaTh2005}.

\begin{figure}[th!]
  \begin{center}
    \subfigure[]{\includegraphics[width=0.45\hsize]{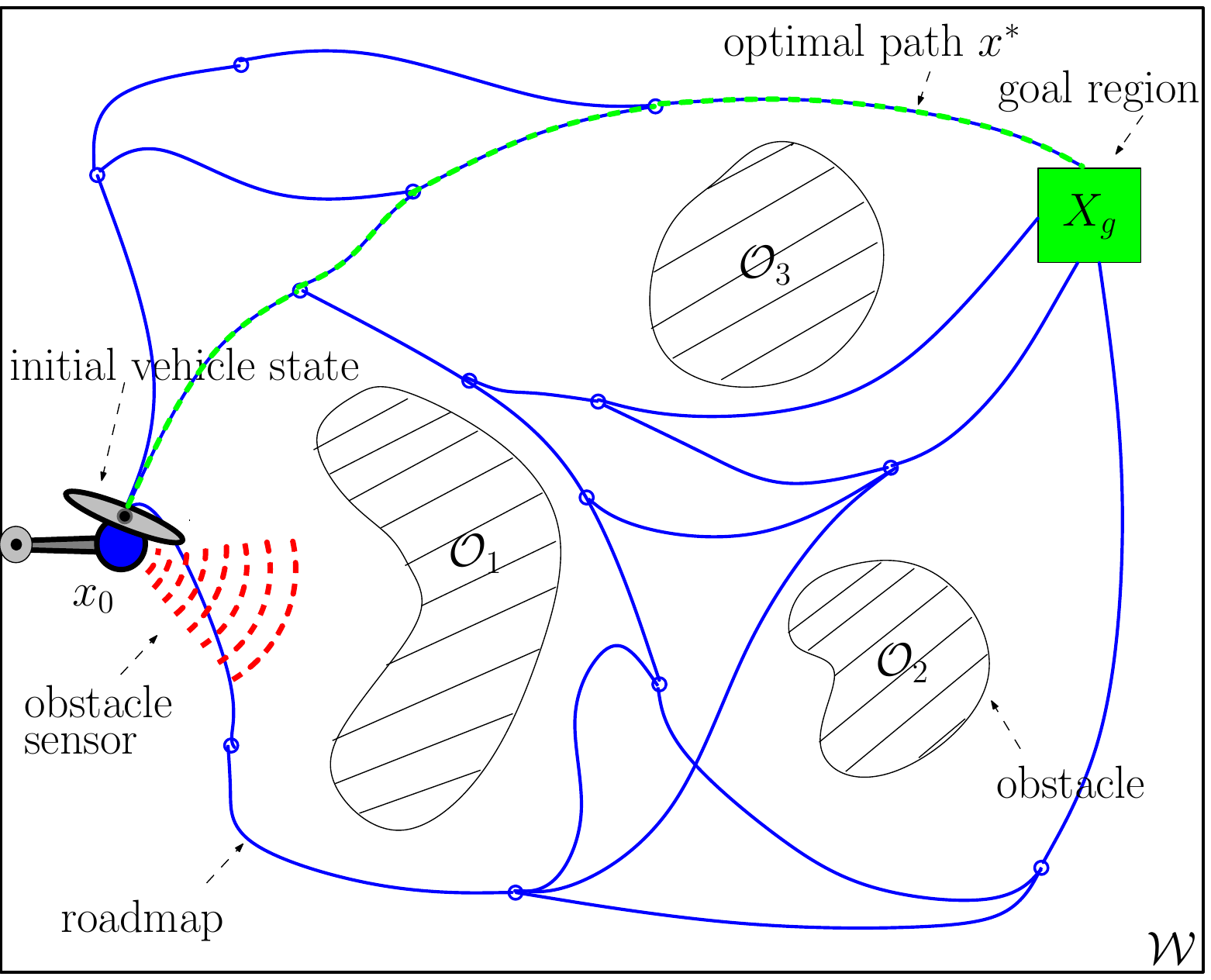}}\hfill
    \subfigure[]{\includegraphics[width=0.45\hsize]{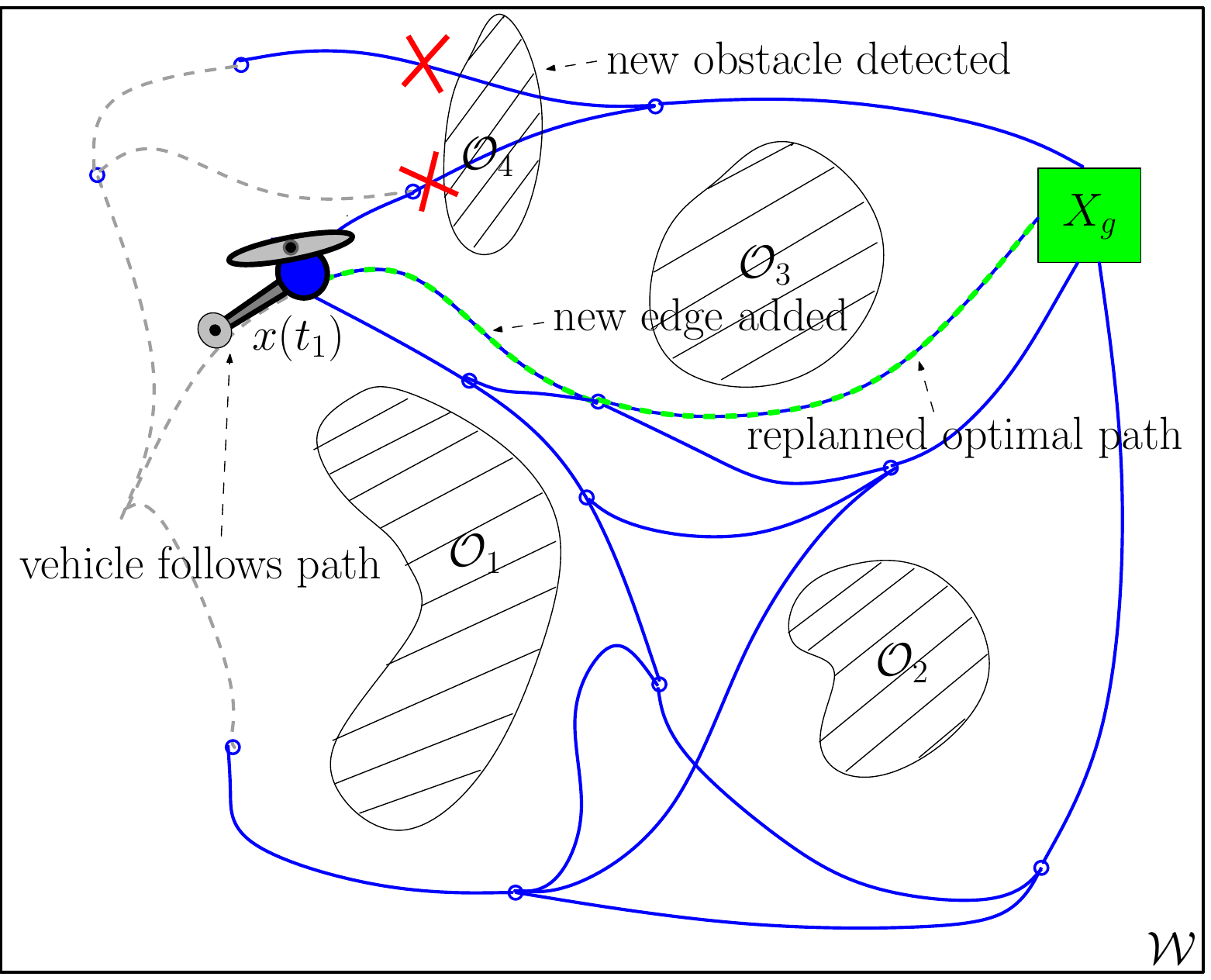}}\hfill\\
    \subfigure[]{\includegraphics[width=0.45\hsize]{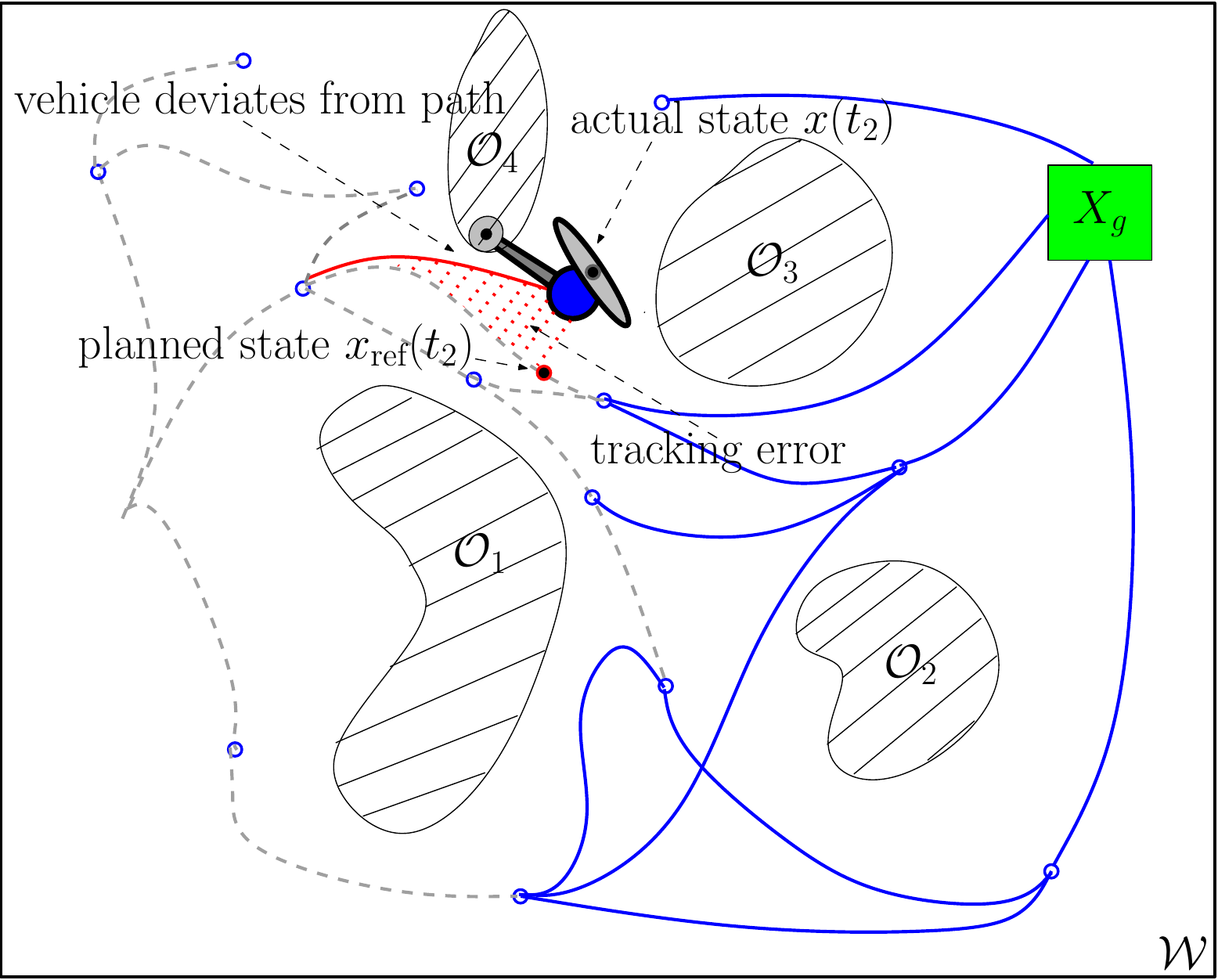}}\hfill
    \subfigure[]{\includegraphics[width=0.45\hsize]{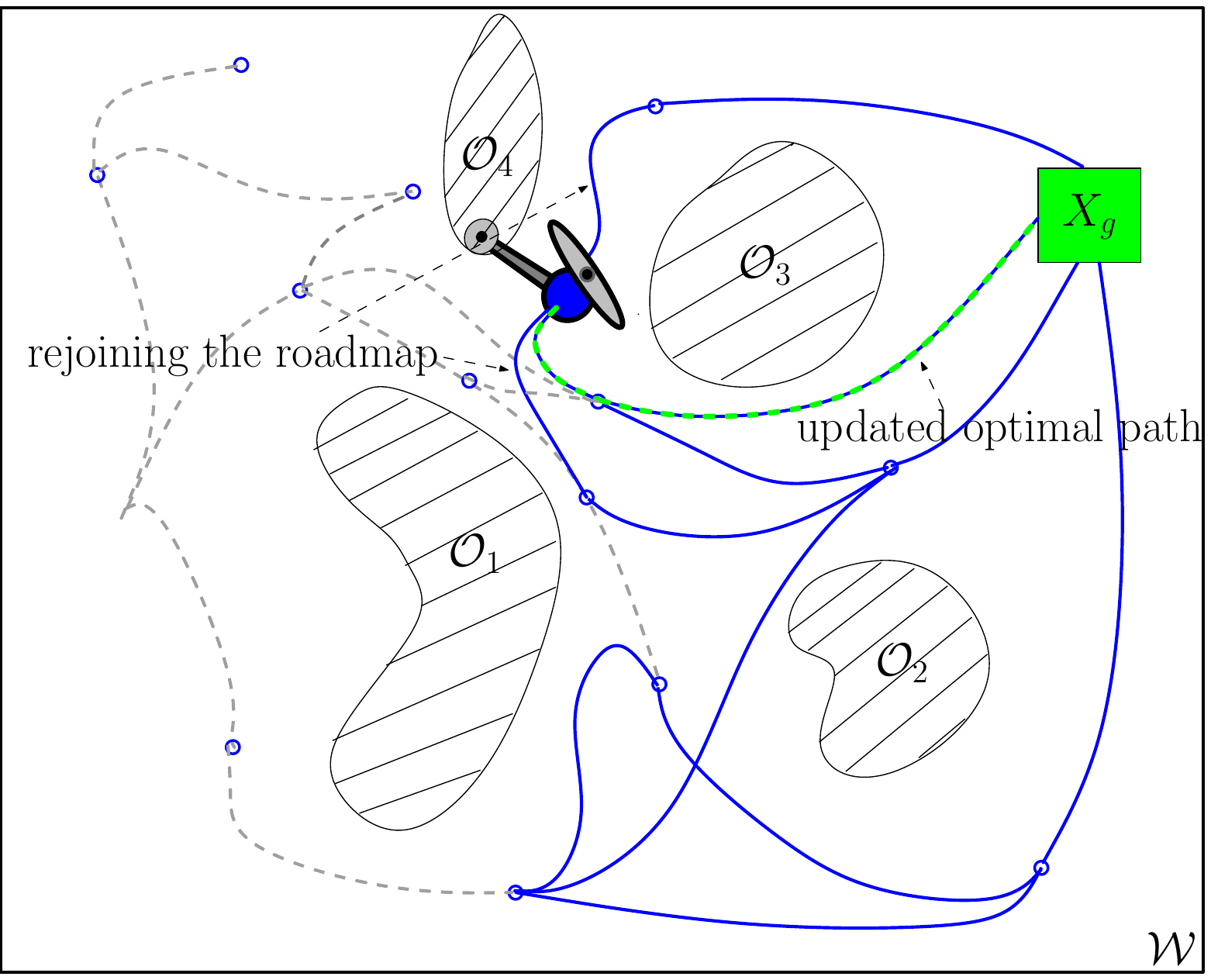}}\hfill
  \end{center}
  \caption{A sketch of a typical planning scenario where a vehicle replans when new obstacles are discovered or vehicle deviates from prescribed planned path. Recomputing a new trajectory in all cases is performed efficiently using $D^*$ search.}\label{fig:prms}
\end{figure}

A key issue in the construction of a PRM is how to connect samples. In
our framework  samples are connected through sequences of locally
optimized motion primitives. The primitives are computed offline and
organized in a library for instant lookup during motion planning. Such
an approach combined with heuristic graph search renders the approach
suitable for real-time applications. Fig.~\ref{fig:prms}(a) illustrates a
constructed roadmap and a computed optimal trajectory.

Note that in this work we employ uniform sampling in order to guarantee complete
exploration of the state space. It should be noted, however, that in many
cases there is problem structure that can be exploited and the nodes
can be chosen in a biased manner, i.e. according to some fitness
function. In general, the issue of \emph{optimal sampling} is complex
and considered an open problem.
% as will be described next.

\subsubsection{Probabilistic Completeness}
Sampling-based methods have been established as an effective
technique for handling (through approximation) motion planning
problems in constrained environments~\cite{La2006}. Since
requiring algorithmic completeness in this setting is computationally
intractable, a relaxed notion of completeness, termed
\emph{probabilistic completeness} has been
adopted~\cite{ChLyHuKaBuKaTh2005} signifying that if a solution exists
then the probability that the algorithm would fail to find it
approaches zero as the number of iterations $n$ grows to infinity.
In addition to being probabilistically complete our approach employs
branch-and-bound and pruning techniques to speedup convergence towards
an optimum.

\subsubsection{Replanning}
The proposed algorithm can naturally handle unmapped
obstacles and recover from trajectory tracking failures. This is
accomplished by removing obstructed edges in the roadmap or adding new
edges from the current state and performing efficient graph
\emph{replanning}.
Fig.~\ref{fig:prms} sketches a prototypical planning scenario in which the vehicle
must plan an optimal route from an initial state $x_0$ to a goal
region $X_g$ in an obstacle-rich environment
with only partial prior knowledge of the environment .
Fig.~\ref{fig:prms}~a) shows initial roadmap construction based on the
prior environment information (e.g. obstacles $\mathcal O_{1,2,3}$).
Trajectory regeneration is utilized to plan around
locally sensed, unknown obstacles obstructing the vehicle's initial
path ($\mathcal O_4$ in Fig..~\ref{fig:prms}~b) ;
imperfect vehicle tracking (Fig.~\ref{fig:prms}~c) of the planned path
is addressed by the insertion of new roadmap nodes matching the
current vehicle state(Fig.~\ref{fig:prms}~d); node insertion is
activated when the error resulting from sensing and actuation noise,
and environmental disturbances exceeds vehicle tracking capabilities.

\subsection{UAV Application}\label{sec:heli}

\subsubsection{Model}
We model the vehicle (a small autonomous helicopter)
as a single underactuated rigid body with configuration space $\mathcal{C}=\operatorname{SE}(3)$
described by position $x \in \mathbb{R}^3$ and orientation matrix $R
\in \operatorname{SO}(3)$. Its {\it body-fixed} angular and linear
velocities are denoted by $\omega \in
\mathbb{R}^3$ and $v \in \mathbb{R}^3$, respectively.
The vehicle has mass $m$ and principal moments of rotational inertia
$J_1,J_2,J_3$ forming the inertia tensor
$\operatorname{\mathbb{J}}=\operatorname{diag}(J_1,J_2,J_3)$.

The vehicle is controlled through a
{\it collective} $u_c$ (lift produced by the main rotor) and a {\it
  yaw} $u_{\psi}$ (force produced by the rear rotor), while the
direction of the lift is controlled by tilting the main blades forward
or backward through a {\it pitch} $\gamma_p$ and sideways through a
{\it roll} $\gamma_r$. The four control inputs then consist of the two
forces $u=(u_c, u_{\psi})$ and the two shape variables
$\gamma=(\gamma_p,\gamma_r)$.

The equations of motion have the standard  form (e.g. ~\cite{BuLe2004}):
\begin{align}
  \left[\!\!\begin{array}{c}
    \dot R \\ \dot x \end{array}\!\!\right] =&
  \left[\!\!\begin{array}{c} R \; \widehat\omega
    \\ R \; v \end{array}\!\!\right], \label{eq:rec} \\
  \begin{split}
    \left[\!\!\begin{array}{c}
      \operatorname{\mathbb{J}} \dot \omega \\
      m \dot v \end{array}\!\!\right] =&
    \left[\!\!\begin{array}{c}
      \operatorname{\mathbb{J}}\omega \times \omega \\
      m v\times \omega + R^T(0,0,-9.81m)\end{array}\!\!\right] +
    F(\gamma)u, \label{eq:ep}
  \end{split}
\end{align}
where the map $\widehat{\cdot}:\mathbb{R}^3
\rightarrow\mathfrak{so}(3)$ is defined by
\[\widehat\omega= \left[\begin{array}{ccc}
    0 & -\omega^3 & \omega^2 \\
    \omega^3 & 0 & -\omega^1 \\
    -\omega^2 & \omega^1 & 0
  \end{array}\right],\]
while the control matrix is defined as
\begin{align*}
  F(\gamma)=\left[\begin{array}{cc}
      d_t \sin\gamma_r & 0 \\
      d_t \sin\gamma_p \cos\gamma_r & 0 \\
      0 & d_r \\
      \sin\gamma_p \cos\gamma_r & 0 \\
      -\sin\gamma_r & -1 \\
      \cos\gamma_p \cos\gamma_r & 0
    \end{array} \right].
\end{align*}
The motion along the trajectories studied next satisfies the dynamics
given in ~\eqref{eq:rec} and~\eqref{eq:ep}.

\subsubsection{Helicopter Primitives}\label{sec:prims}
The local motion planning method used in the framework developed in this paper is
based on computing a sequence of \emph{motion primitives} that exactly
satisfy the boundary conditions, i.e., exactly reaches a sampled
node. The symmetry in the system dynamics allows us to employ a
\emph{maneuver automaton} to produce sequences of continuously
parametrizable motions (trim primitives) connected with
maneuvers. This general framework developed in~\cite{FrDaFe2005}
is suitable to systems such as UAVs or ground robots if one ignores
pose-dependent external forces, such as varying wind or changing
ground friction as function of position.

Let the vehicle rotation be described by its roll $\phi $,
pitch $\theta$, and yaw $\psi$. Denote the linear velocity by $
v=(v_x,v_y,v_z) \in \mathbb{R}^3 $, and the angular velocity by $
\omega=(\omega_x,\omega_y, \omega_z) \in \mathbb{R}^3 $.
Denote the whole configuration by $ g \in SE(3) $, the whole velocity by
$ \xi_b \in \mathfrak{se}(3)$, defined by
\begin{align}
  g = \left[\begin{array}{cc}
      R & p \\
      \mathbf{0} & 1
    \end{array}\right], \quad
  \xi_b = \left[ \begin{array}{cc}
      \widehat\omega & v \\
      \mathbf{0} & 0
    \end{array}\right].
\end{align}

By defining the map
$\mathbb{I}=\operatorname{diag}(J_1,J_2,J_3,m,m,m)$ the dynamics can
be expressed in more general form as
\begin{align*}
\mathbb{I}\dot\xi_b =  \operatorname{ad}_{\xi_b}^* \operatorname{\mathbb{I}} \xi_b + f_u + \operatorname{Ad}^{\ast}_{g} f_{ext},
\end{align*}
where $f_u$ is the control force, corresponding to $F(\gamma)u$
in~\eqref{eq:ep}, while $f_{ext}= (0,0,0,0,0,-9.81 m) \in
\mathfrak{se}(3)^* $ is the gravity force. Since gravity is the only
configuration-dependent term in the dynamics and is
invariant to translations and rotations the $ z $-axis, then the
dynamics symmetry group can be set as $ G = SE(2)\times \mathbb{R}$.

The motion planning problem is solved in closed form through inverse
kinematics of a minimal number of primitives (total of five).
For a full description of the trim velocity invariance conditions, the construction of maneuvers
between two trim motions, and all other details of the design of the trims and maneuvers
we refer the reader to \cite{Ko2008}.

The continuous optimal control formulation associated with the maneuver determination
is computationally solved
through the discrete mechanics methodology~\cite{MaWe2001} which is
particularly suitable for systems with nonlinear state spaces and
symmetries. A geometric structure preserving optimizer developed
in~\cite{Ko2008} (Section 2.7) was used to perform the
computations. The computations were performed offline with the
resulting optimal maneuvers assembled in a library offering instant
lookup during run-time.

\subsubsection{Test scenarios}
\subsubsection*{Mapping}
The terrain is represented using a digital elevation map loaded
from a file. The obstacle proximity function ${\bf prox}$
(see~\eqref{eq:prox}) between the helicopter trajectory and the
terrain is computed using the Proximity Query Package (PQP)
\cite{GoLiMa1996} that can compute closest distance between two
arbitrary meshes.

\subsubsection*{Simulation Analysis}
The PRM algorithm is tested through simulation. Its performance is
evaluated as a function of two parameters: the number of nodes and the
environment difficulty. The recorded metrics are: (1) the CPU runtime of
(i) off-line construction, (ii) start/goal setup, and
(iii) planning (i.e. dynamic programming/graph search), and (2) the trajectory cost.

\begin{figure*}[t!]
  \begin{center}
    \subfigure[]{\includegraphics[width=0.4\hsize]{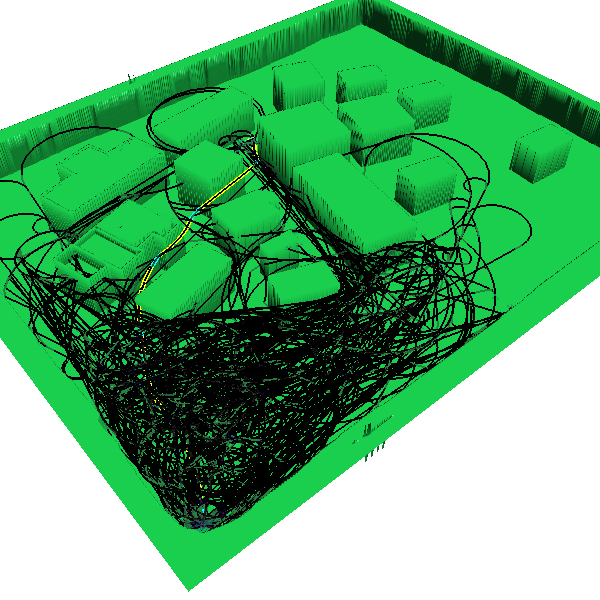}}\\
    \subfigure[]{\includegraphics[width=0.4\hsize]{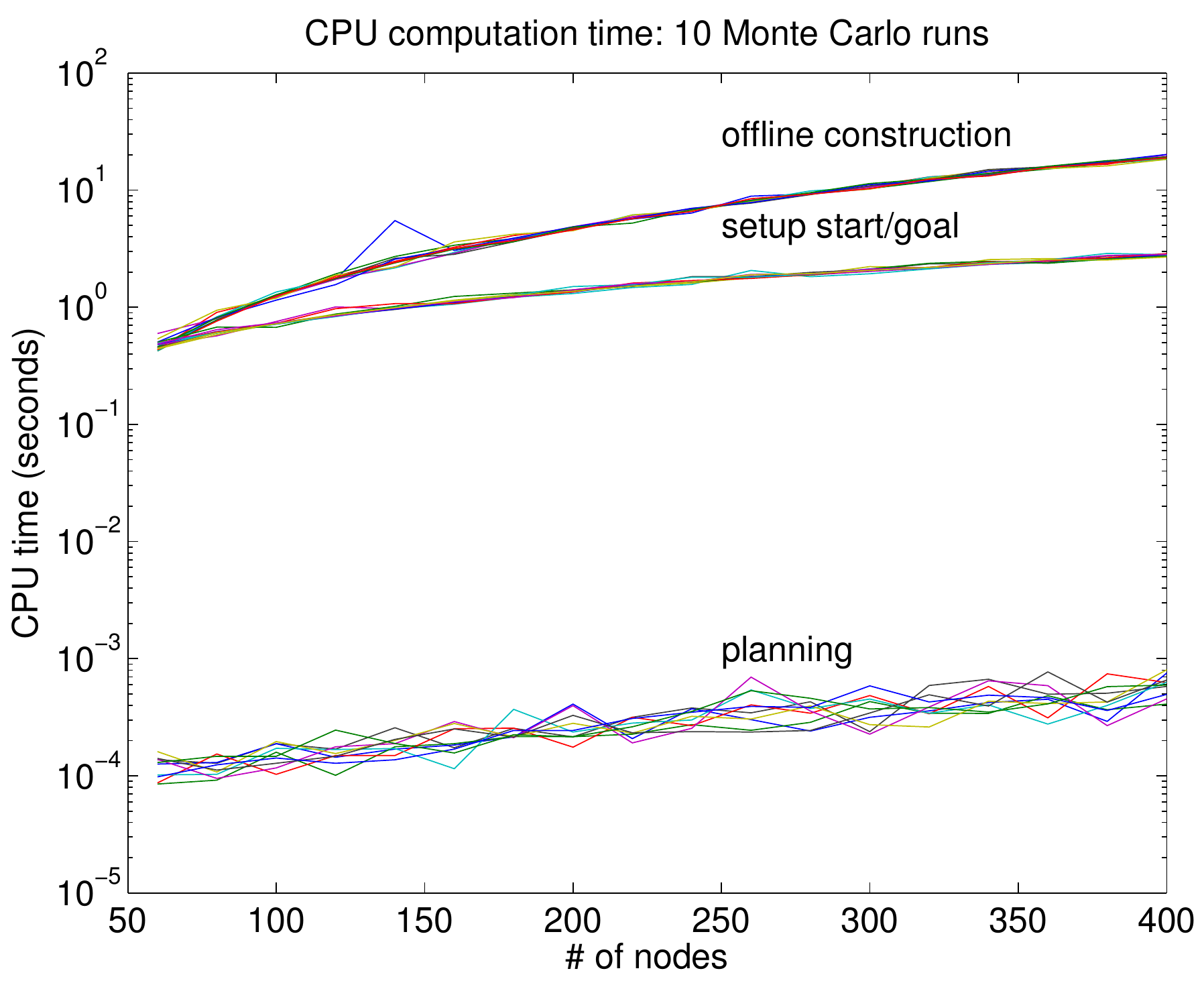}}\hspace*{2cm}
    \subfigure[]{\includegraphics[width=0.4\hsize]{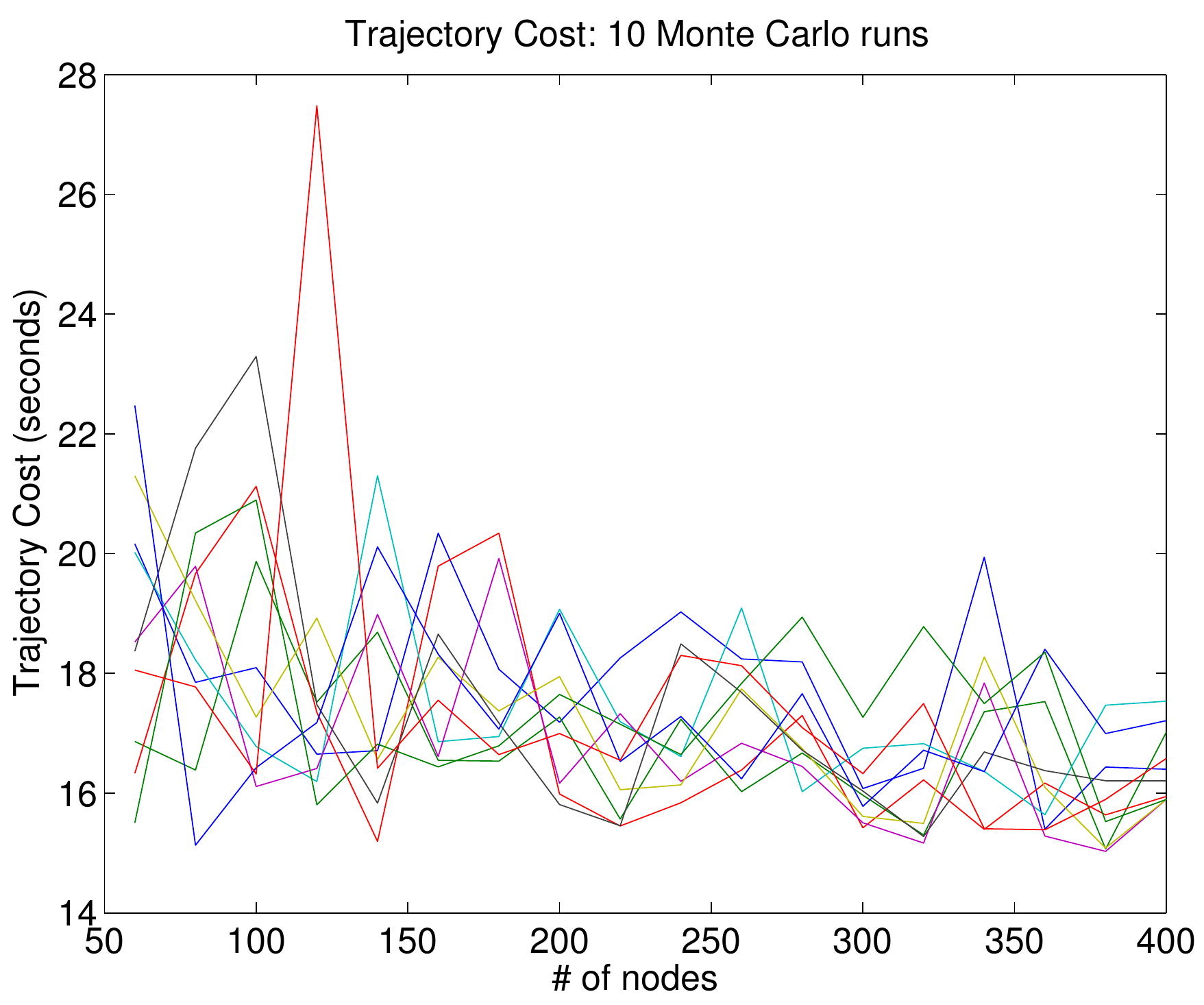}}
  \end{center}
  \caption{(a) part of the roadmap; (b) initial path; (c)
    update map and replan}
  \label{fig:log1}
\end{figure*}

\paragraph{Number of Nodes}
The range of 60-400 nodes is tested. A roadmap is constructed covering
the state space without specified start and goal (step1: off-line
construction). Once it is finished, start and goal are specified and
are connected to all possible nodes (step2: start/goal setup). Then a
path is planned (step3: graph search). Fig.~\ref{fig:log1} shows that
the cost of optimal trajectory improves as a function of the nodes in
the graph.

\begin{figure*}[t!]
  \begin{center}
    \subfigure[]{\includegraphics[width=0.4\hsize]{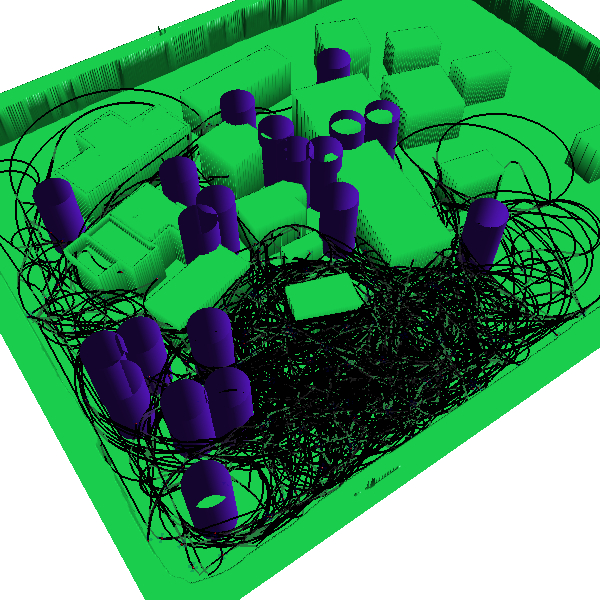}}\\
    \subfigure[]{\includegraphics[width=0.4\hsize]{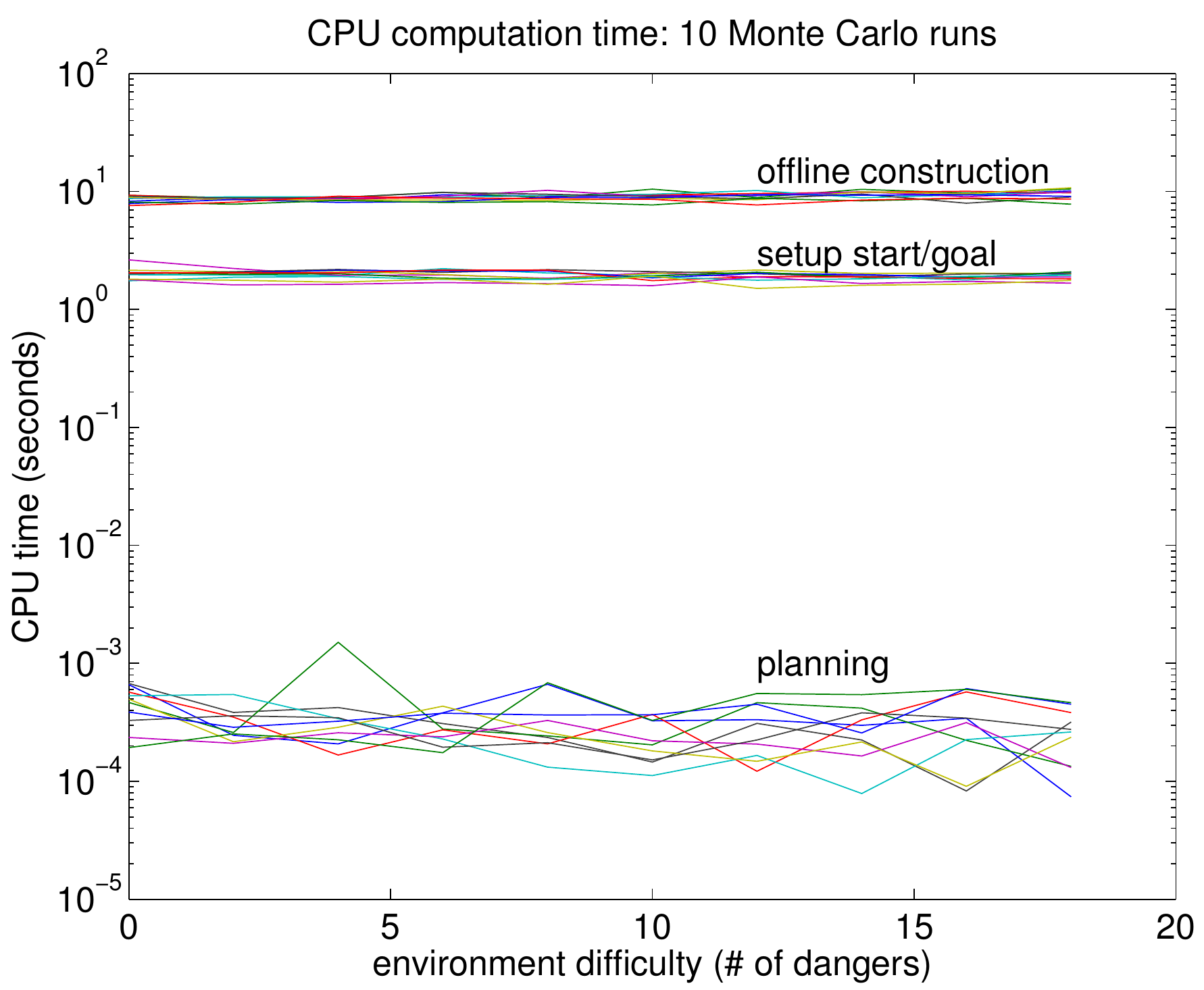}}\hspace*{2cm}
    \subfigure[]{\includegraphics[width=0.4\hsize]{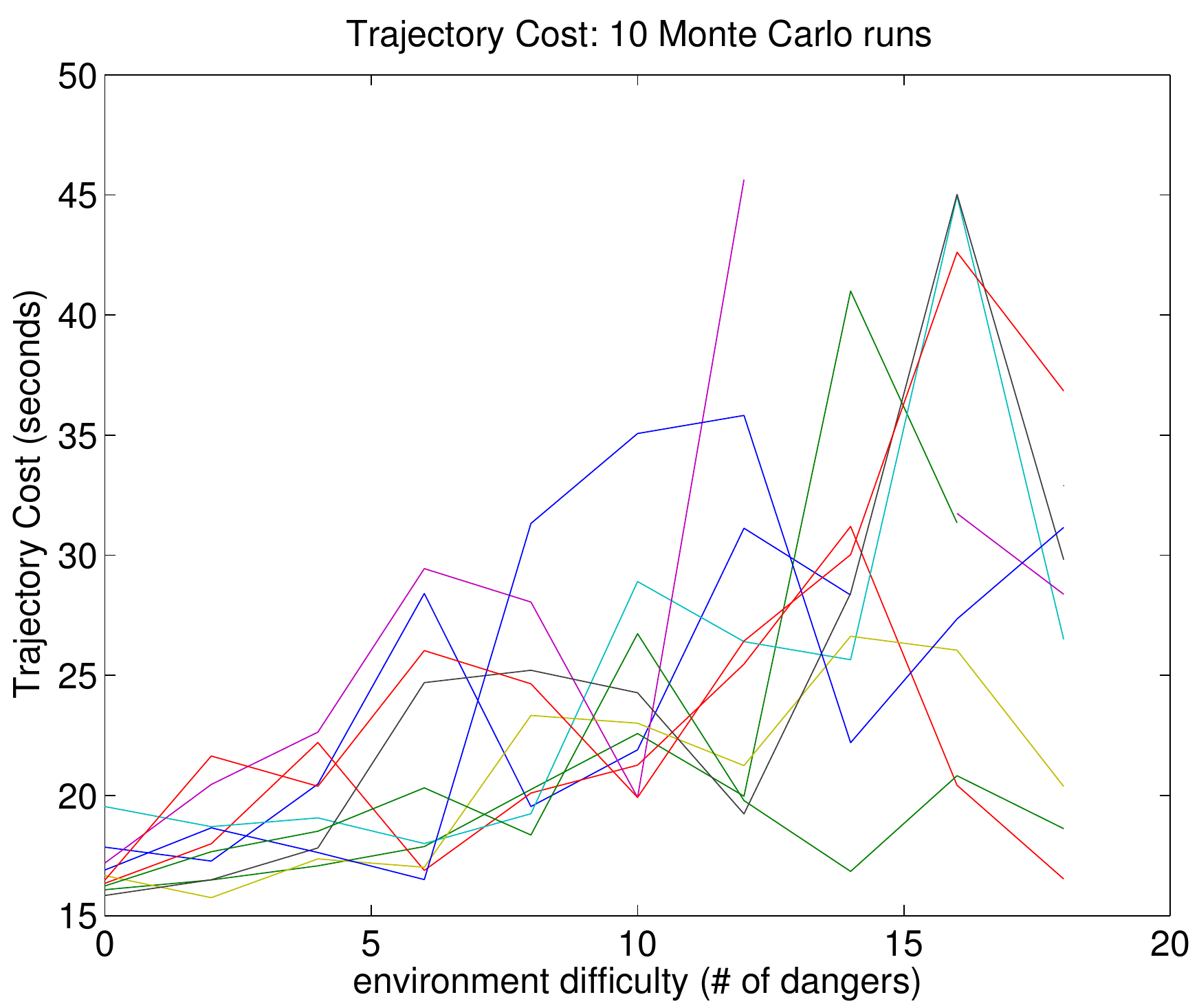}}
  \end{center}
  \caption{(a) part of the roadmap; (b) initial path; (c)
    update map and replan}
  \label{fig:log2}
\end{figure*}

\paragraph{Environment Difficulty}
The basic Fort Benning (FB) environment is populated with cylindrical
obstacles (threats) the number of which determines the environment
difficulty. The threat locations are randomly chosen along a trajectory
computed in the previous cycle (i.e. in the "easier"
environment) thus making the planning task progressively more
difficult. Each run employs 250 samples with the results shown on
Fig.~\ref{fig:log2}. As expected, the trajectory cost grows as the
environment becomes more cluttered. The computation does not change
substantially.
.

Our simulation tests have been performed using uniform, unbiased
sampling. While this is not the most computationally efficient
strategy it is provably convergent. Efficiency can be additionally
improved through smarter sampling, e.g. along the medial axis, and
through a more efficient ways to ``join'' the graph during
replanning. It should be noted that the planning step is still
super fast: even as the environment becomes very cluttered graph
search requires only a fraction of a millisecond (Fig.\ref{fig:log1}
b) and Fig.\ref{fig:log2} b)).

\subsubsection*{Experimental Tests}
\begin{figure*}[t!]
  \begin{center}
  \subfigure[]{\includegraphics[width=0.65\hsize]{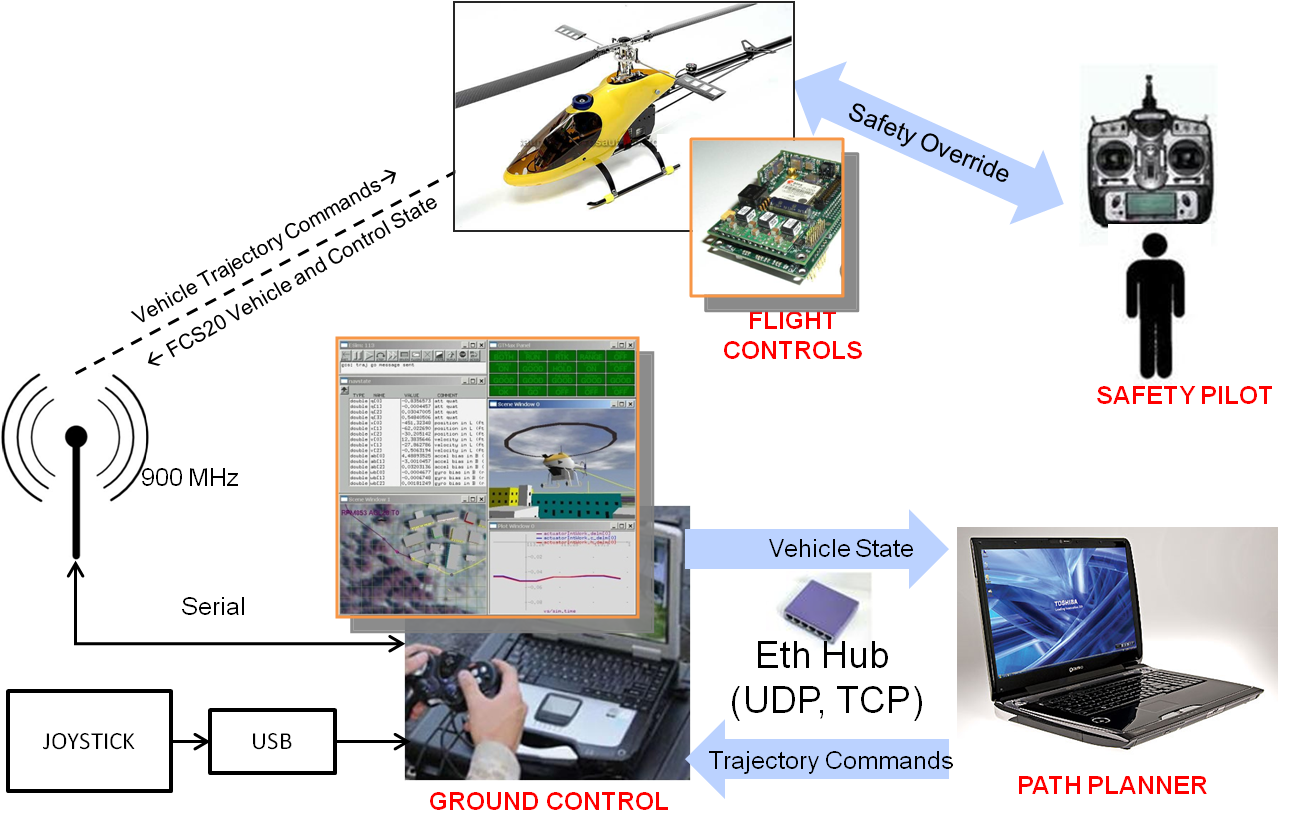}}\\
  \subfigure[]{\includegraphics[width=0.65\hsize]{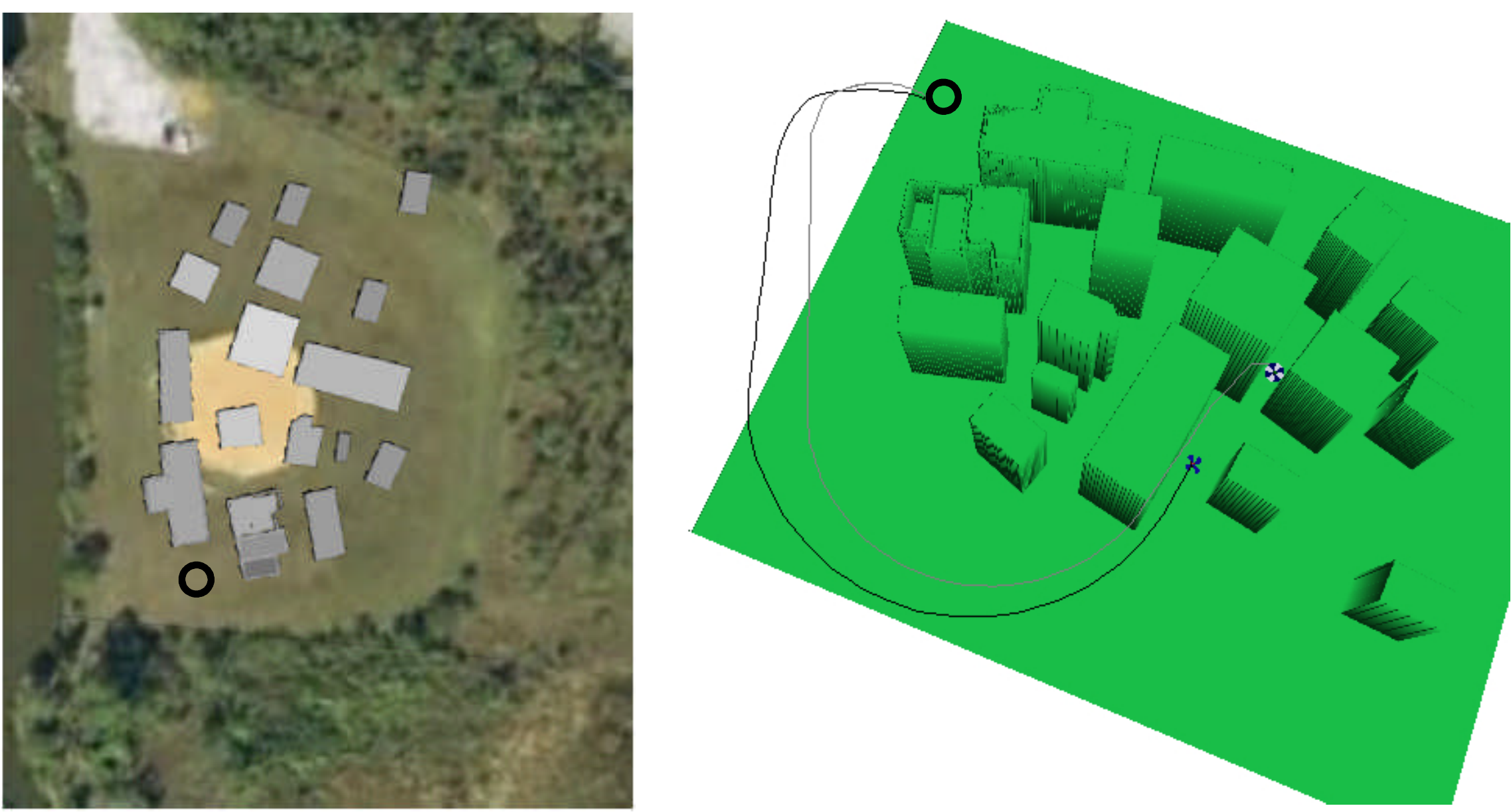}}
  \end{center}
  \caption{a) Schematic of the experimental setup. Left picture in Fig b) shows the virtual obstacle set (gray rectangles) superimposed on West Palm Beach
   experimental test site (test flight start location indicated as open black circle); right) experimentally commanded (gray curve) and followed (black curve) vehicle paths around virtual obstacles (test flight start location indicated as open black circle).}
  \label{fig:expt0}
\end{figure*}

We have recently completed experimental flight tests (see Fig.~\ref{fig:expt0}a) for the experimental setup) for a maxi-joker UAV planning using the PRM approach described. Fig.~\ref{fig:expt0} b) (left) shows an aerial photo of the experimental test site (West Palm Beach, FL) with a collection  of \emph{virtual} obstacles (known to the planner), from the Fort Benning map, superimposed. The PRM approach is used for real-time trajectory planning from a start state (indicated by black open circles in Fig.~\ref{fig:expt0}) b) to a goal state amongst the virtual obstacles. Our experiments successfully demonstrated the real-life application of PRM to trajectory planning (see followed trajectory in Fig.~\ref{fig:expt0}(right)) in a cluttered, obstacle-rich environment using realistic UAV flight dynamics.

.

\section{Design Methodology for Uncertain Systems}\label{sec:Design flows}
In previous sections we have presented techniques for uncertainty
quantification with applications to analysis and synthesis. In this
section we discuss the general problem of organizing a design flow for
complex systems in the face of uncertainty.

\begin{figure}[t!]
  \centering
  \includegraphics[width=0.70\hsize]{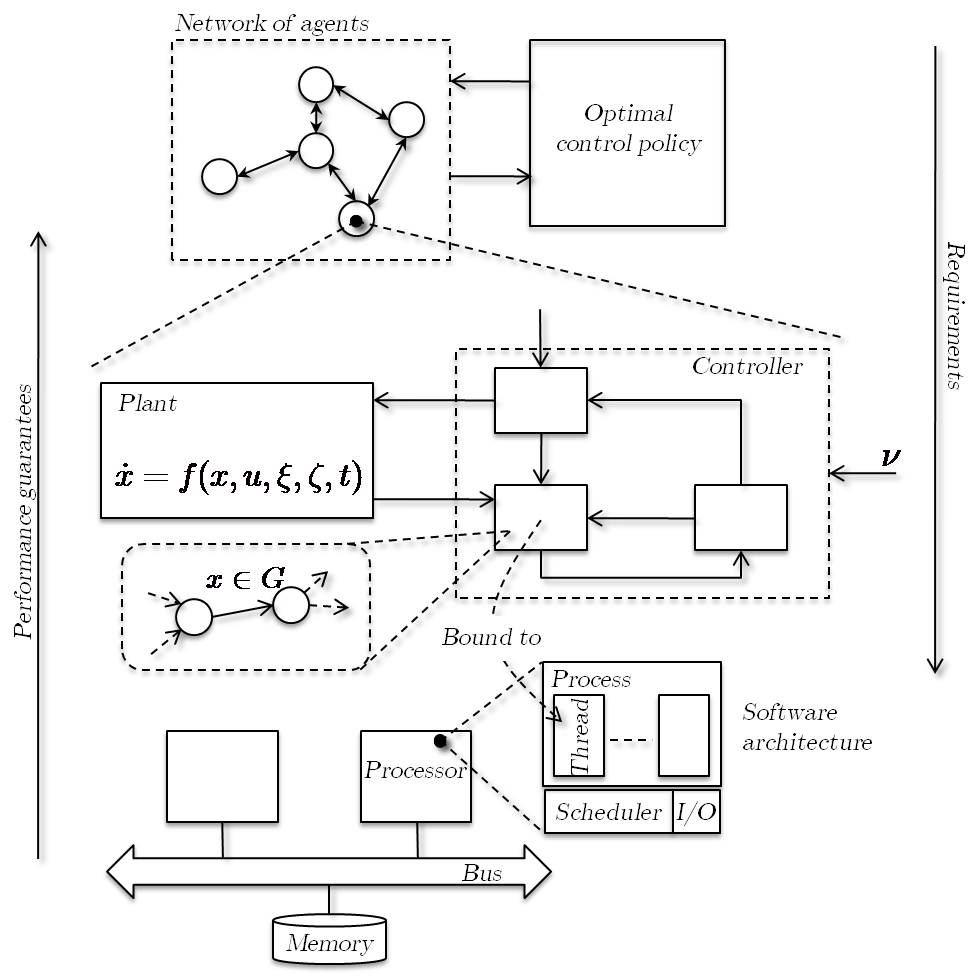}
  \caption{High level view of the organization of a design flow including
  sources of uncertainty.}
  \label{fig:anatomy-spec}
\end{figure}

Figure~\ref{fig:anatomy-spec} is purposely divided into two parts. The upper
part shows a network of agents and their dependencies (represented by double
pointed arrows). As shown in Section~\ref{sec:algorithm}, given the goal of
searching an area for targets, and optimal control strategy can be computed that
coordinates the vehicles to achieve coverage. The target initial position and
motion, as well as the performance of the agents can be considered uncertain and
the controller needs to take this uncertainty into account.

The same problem repeats at the agent level. We use the bottom part of
Figure~\ref{fig:anatomy-spec} as a general representation of the sources of
uncertainty that appear in a model. The figure shows the internal details of an
agent. The agent (referenced to as {\em plant}) is characterized by its
dynamics that is affected by noise (a
stochastic process $\zeta(t)$) and by uncertainty in some of its
parameters (the vector of random variables $\xi$). At this abstraction
level, the digital controller is considered ideal, meaning that
resource constraints are not embedded in the model. This abstraction
removes part of the uncertainty coming from performance metrics and
failures. However, transitions in the controller may be guarded by
expressions that make explicit reference to the state of the plant ($x
\in G$ in Figure~\ref{fig:anatomy-spec}) which is a random process,
hence transitions are taken with some probability. Most importantly, some
control algorithms are randomized, meaning that their execution depends on a
vector of random variables $\nu$ that make the execution different at each
invocation of the algorithm. An example of randomized algorithm is the one
presented in Section~\ref{sec:Robust Path Planning} where a sampling technique
is used to sample the trajectory space. As a result, the performance of a
control algorithm can only be characterized in probabilistic terms.

Control algorithms are ultimately executed on a possibly distributed
architecture. The architecture comprises the description of the hardware (which
includes processors, storage elements, communication networks and
interconnections among them), and the software (which includes processes,
threads, schedulers and I/O interfaces). An architectural component may be
characterized by a behavioral model as well. For example, the scheduler used by
the real-time operating system or by the communication protocol is implemented
by a state machine. Moreover, architectural components are annotated with
performance metrics such as delay and failure rate.
Performance metrics are usually probabilistic. Optimal control
strategies rely on the solution of optimization problems whose
run-time depends on the input data. Moreover, the worst case execution
time of software is data dependent because of low level implementation
techniques such as cache memories, branch prediction, pipeline
execution etc. Notoriously, communication delays are also uncertain,
especially when collision-based and wireless protocols are used.

It is clear from this simple yet comprehensive description that systems are
{\em complex}, namely involving a large number of components and interactions,
{\em heterogeneous}, namely comprising continuous dynamics as well as discrete
transition systems, and {\em uncertain}. The techniques presented in
Section~\ref{Bottom up graph decomposition} and~\ref{Scalable Uncertainty
Quantification} are general tools that represent a major
advancement in terms of the size of systems that can be analyzed. The key idea
in these techniques is to partition the system into loosely coupled sub-systems.
A design flow that is based on a vertical decomposition of the problem into
abstraction layers can push even further the complexity wall. The key idea is to
engineer a design flow where sub-systems are abstracted into performance
models to be used at higher abstraction levels (e.g. the dynamics and the
controller of an agent are abstracted into a new and simpler dynamical model of
an agent to be used in a network of agents). At the same time, the result of a design choice
at one level is flown down to sub-systems (e.g. the agent must respond to a
trajectory command within a given bounded time, and the maximum deviation
of the actual trajectory from the required trajectory needs to be within given bounds).

\subsection{Overcoming complexity}
\label{sec:complexity-abstractions-tools}
Figure~\ref{fig:anatomy-spec} gives the idea of the complexity of these systems
when all details are considered together. Despite the efficiency of
uncertainty quantification techniques, such complexity is out of
reach. Abstraction from details is key to overcome complexity.
Another effective way is to avoid undertaking complex analysis tasks by {\em
synthesizing} part of the system so that some properties are guaranteed (and
therefore do not need to be checked). A synthesis algorithms is given a partial
description of the system and a performance goal, or in general some properties
that must hold true. The algorithm synthesizes a controller that when
connected to the partial model satisfy -- by construction -- the user defined
properties.

\begin{figure}[t!]
\begin{center}
  \includegraphics[width=0.70\hsize]{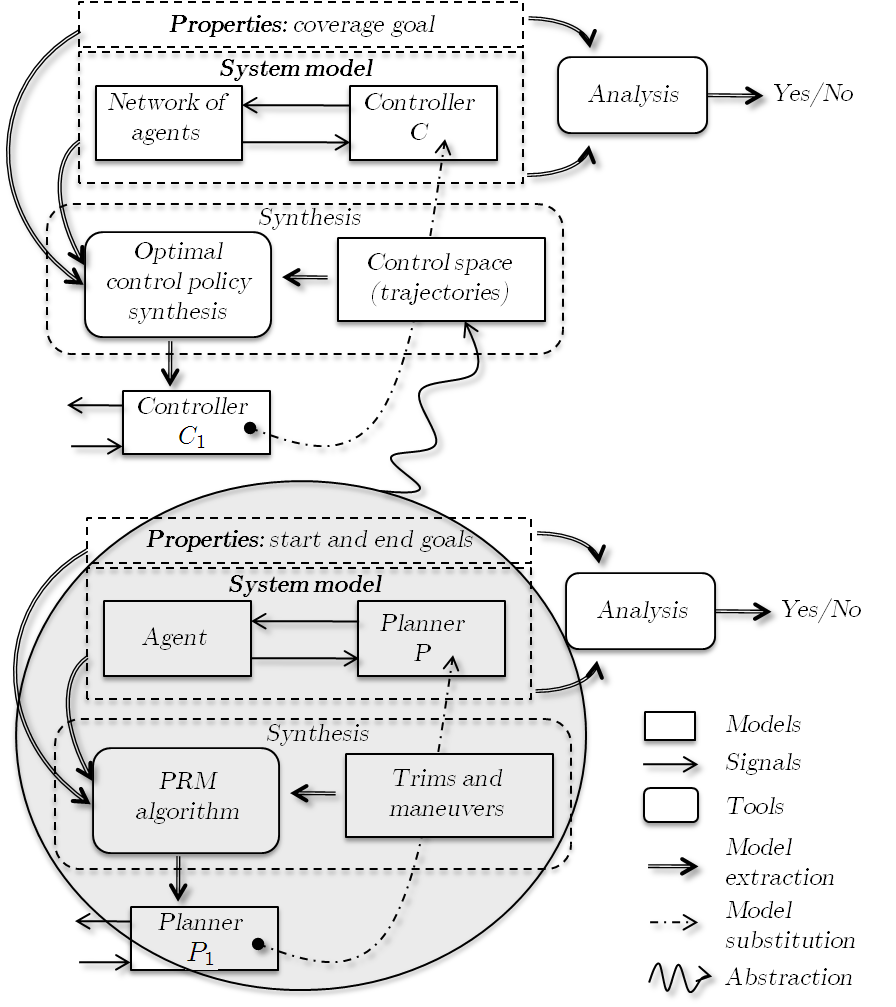} %[width=\columnwidth]
  \caption{Organization of models, tools and abstraction layers.}
  \label{fig:models-tools-abstractions}
\end{center}
\end{figure}

Figure~\ref{fig:models-tools-abstractions} shows the overall design methodology
together with the relationships between models, tools and abstraction layers. We
considered the applications presented in this paper, namely the optimal coverage
problem and trajectory planning. Given models for a network of agents and a
controller, and given some coverage goals (e.g. probability of covering a
certain region), uncertainty quantification can be used to check that the
property is satisfied by the model. An alternative path is synthesis. In this
case, the goal and model are used as optimality criteria and constraints of an
optimization problem that selects a control strategy for the network of agents
among a set of admissible ones. The new controller $C_1$ can be substituted to
the control block in the system and be sure that the coverage goal is satisfied.
In some cases, analysis might still be needed as the synthesis algorithm may be
heuristic. The analysis result could then be used to tune some of the parameters
of the algorithm.

The path planning problem can be cast in the same general framework. The
abstraction level is lower as the real dynamics of the vehicle are considered
and more details are embedded in the description of the environment. The
algorithm presented in Section~\ref{sec:Robust Path Planning} synthesizes a path
for a vehicle by selecting a sequence of trims and maneuvers taken from a
library. In this case, the algorithm is randomized and analysis might be used to
compute the probability that  the vehicle reaches the end goal for a given
distribution of obstacles and parameter values.

Interestingly, one could simplify the optimal coverage problem by relying on a
path planner that provides probabilistic guarantees of finding paths between
points in the presence of known and unknown obstacles. This abstraction is used
as a model of the motion of the vehicle and the available control actions. The
new control space for the optimal control policy generation algorithm is the set
of commands moving the vehicle from a state (including position, velocity and
attitude) to another state (described as a new position, velocity and attitude). The
abstraction also includes a probabilistic guarantee that the command will be
executed. Building this abstraction is non trivial. Interestingly enough, such
abstraction can be build using uncertainty quantification at the lower level.
In Section~\ref{sec:Robust Path Planning}, a detailed simulator is used to
compute the cost of path and the same type of techniques can be used to compute
the probability that a path can be found for a given number of nodes sampled by
the PRM algorithm. This performance metrics are exported to the optimal coverage
algorithm that can therefore plan a trajectory trading off coverage, time, and
probability of mission success.

%\begin{example}[Abstraction of the PRM]
%\end{example}

In Section~\ref{sec:algorithm}, a different abstraction has been used
regarding the vehicle and its control. In fact, the vehicle dynamics has been
assumed to be a first or second order dynamics. This can also be considered an
assumption on the vehicle dynamics that must be respected when designing the
low level control of the vehicle. In our methodology, this assumption becomes a
constraint or requirement that is flown down to the control design problem.

\subsection{Complexity arising from heterogeneity}
\label{sec:complexity-heterogeneity}
Another source of complexity in these systems is the level of heterogeneity in
the models. Some of the models are described in terms of differential equations.
However, there are control algorithms that are discrete and represented by
finite state machines. The PRM algorithm in Section~\ref{sec:Robust Path
Planning} is an example of controller that is described in terms of a complex
sequential program. In other cases, a probabilistic finite transition system
(e.g. a Markov Chain or a Markov Decision Process) is used as abstraction of the
underlying complex system. The abstraction of the PRM performance is a good
example.

Although some techniques (such as graph decomposition) can be used in both cases
(by changing the definition of the graph), there are techniques that work best
on dynamical systems (such as uncertainty quantification), and other techniques
that are more appropriate for probabilistic transition systems (such as
Probabilistic Model Checking~~\cite{Alfaro95modelchecking,Baier1998,prism}, and
performance and reliability analysis using Markov Chains~\cite{qnmc_book}). For
example, the probabilistic analysis of computation and communication
architectures has been traditionally done using Markov Chains (perhaps starting
from higher level descriptions such as Stochastic Petri
  Nets~\cite{gspn} and Stochastic Automata Networks~\cite{Plateau1991}).

The traditional way to overcome the heterogeneity problem is again by
introducing a separation between the control world (higher abstraction levels
and typically continuous time) and the digital world (lower abstraction levels
and typically discrete). Once a control algorithms has been
  designed and analyzed, constraints such as maximum communication delay and
  maximum jitter can be derived for the underlying architecture. The
  architecture design problem then is to find a cost effective architecture that
  satisfies a constraints on the delay distribution.

Unfortunately, it is not always possible to
clearly separate continuous systems and finite state systems for several
reasons. In some cases, the dynamics of the plant is hybrid as result of an
abstraction process where very fast but irrelevant dynamics are neglected and
lumped into instantaneous jumps. In other cases, the controller is simply
discrete, or the time scales of the control algorithm and of
the underlying platform are close, hence one has to considered the effect of
software transitions on the dynamics of the system. In general,
the system is a stochastic hybrid system (SHS)~\cite{Hu_shs_hscc2000}.
Analysis and design tools for SHSs are still in their infancy but some
promising results have started to appear in
literature~\cite{AKLP10,ShsMathewAllerton2010,Koutsoukos_computationalmethods}.

\subsection{Tooling}
\label{sec:methodology-tooling}
Figure~\ref{fig:models-tools-abstractions} shows some interesting aspect of a
design flow: the organization of models and tools. From the previous sections,
it is clear that analysis and synthesis algorithms require their inputs to be
captured in a precise mathematical description. Thus, the use of formal
languages is a key enabler to implement a design flow. Once formal languages
for continuous time systems and finite state systems are available, models can
be stored in a persistent format, and loaded into a data structure in memory on
a computing machine. However, each tool might require inputs in a specific format
and may provide outputs also in a specific format. Thus, model transformation
technologies need to be used to ensure a uniform representation in a model
database. Mode transformation (in some cases also referred to as model
extraction) is a non-trivial problem as will be appear clear in
Section~\ref{sec:design-flow-examples}. Finally, efficient tools need to be
developed and linked into a design flow.

While these key elements of a design flow have reached a certain level of
maturity for non-uncertain systems (e.g. the use of high level
languages based on models-of-computation, model-checking,
schedulability analysis, logic synthesis, and mapping), the
level of maturity for probabilistic systems is very low.

At the language level, a system designer should be allowed to define
precisely the source of uncertainty which can include inputs to the
system driven by stochastic processes, or parameters whose values are
characterized by certain probability distributions. Designers
routinely include these type of information in their high level
descriptions using random number generators. These information are
mainly used in simulation and not much in analysis and optimal
design. The reasons cannot be attributed to the lack of tools. In
fact, although in their infancy, formal methods for probabilistic
systems are available (see for example Probabilistic Model
Checking~\cite{prism}, Fault Tree Analysis~\cite{fault_tree_nasa},
reliability and performance analysis using Markov
Chains~\cite{qnmc_book}, uncertainty quantification~\cite{sde,gPC}).

We believe that the challenges in embracing uncertainty in the design
of cyber-physical systems are of different nature and include the sheer
computational complexity of many analysis and optimization problems dealing
with randomness. Specifically, we believe that the challenges are:
\begin{itemize}
\item Model-based design tools used in industry such as Simulink/Stateflow provide high level languages for capturing the design specification. On the other hand, analysis and synthesis tools for probabilistic systems accept specifications described using languages at a much lower abstraction level such as Stochastic Petri Nets ~\cite{gspn}, Stochastic Automata Networks ~\cite{Plateau1991} or directly Markov Chains. The translation from high level languages to low level ones is also referred to as the model extraction problem which is a complex task to automate.
\item Fast analysis methods for heterogeneous uncertain systems including (but
not limited to) Stochastic Hybrid Systems. Ideally, these methods should be able
to include other models such as data-flow models. Extension of graph decomposition (Section \ref{Bottom up graph decomposition}) and PWR approach (Section \ref{Scalable Uncertainty Quantification}) to deal with such systems is also a challenge.
\item Ability to present result to the user of these tools. It is often the case
that analysis techniques operate on a transformed version of the state space
and that properties are defined on computation paths rather than on single
states. In this cases, providing feedback to the user in terms of the reasons
why a property is violated is non-trivial.
\end{itemize}

\subsection{Building a design flow for uncertain systems}
\label{sec:design-flow-examples}
A tool to analyze and design these type of systems must be independent
from the input model and should only be based on the assumptions that
can be made about the modeling languages used to capture the
specification. This tool should accept a functional model described as
a stochastic hybrid system, an architectural
model including performance annotations, and the specification of the
binding of the functionality (i.e. the controller) on the
architectural resources (i.e. processors, networks and storage
elements). Designers should be given the opportunity to define
parametric uncertainties in the input model, namely symbolic
variables representing transition probabilities, that can be used to
sweep over a range of possible values in the performance analysis
step. Because of the complexity of the system description, the result
of the analysis step is typically difficult to interpret. Thus,
designers should also be provided with a practical way of getting
insightful information from the result of the analysis.
%In particular, it should be possible to
%project the proabilisitc result onto a restricted set of signals and
%states.

\begin{figure}[t!]
  \centering
  \includegraphics[width=0.70\hsize]{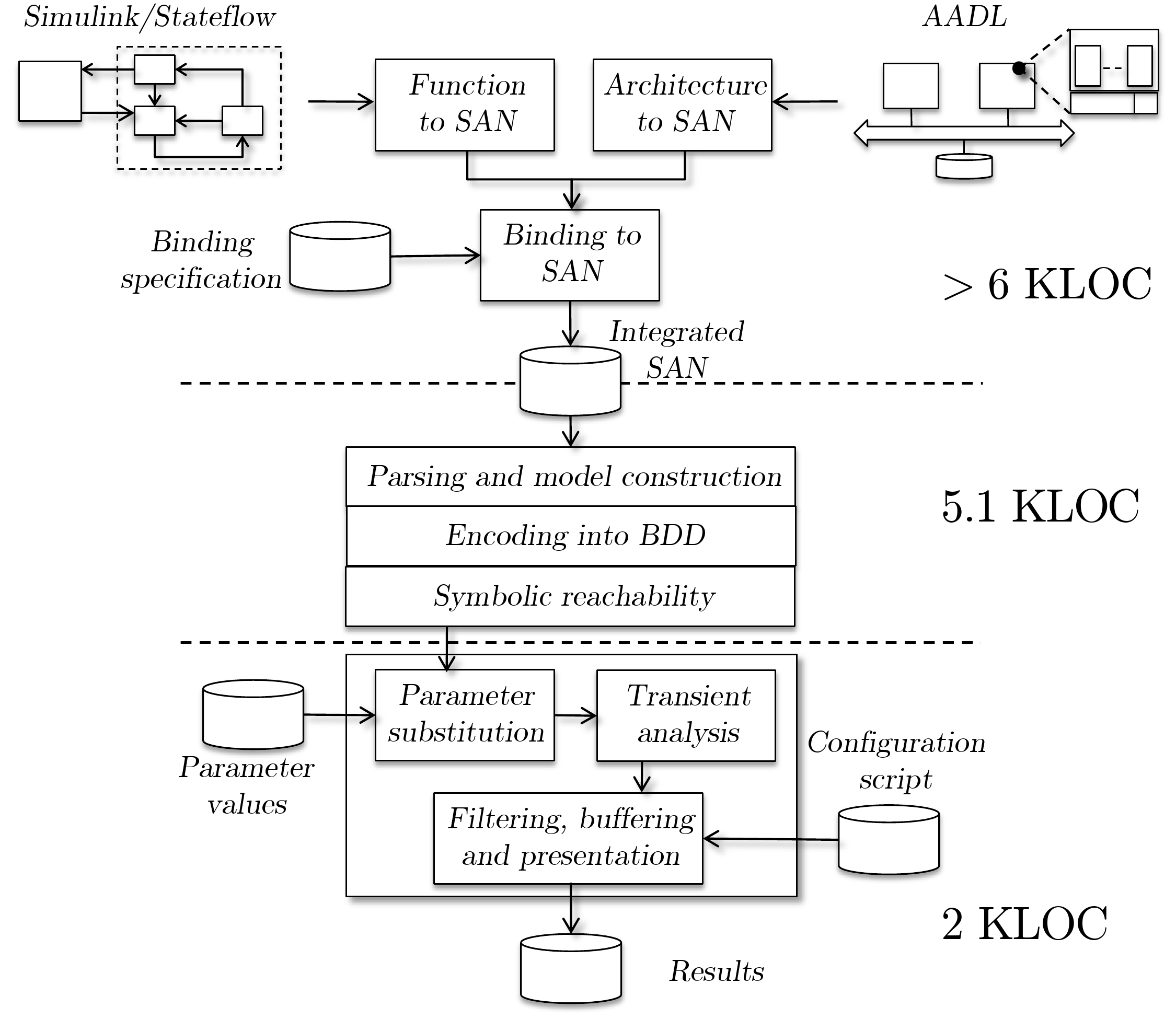}
  \caption{Description of the design tool for uncertain systems.}
  \label{fig:anatomy-tool}
\end{figure}

Figure~\ref{fig:anatomy-tool} shows the architecture of the tool we
developed. To provide all the
aforementioned features, the tool is divided into two parts, a front-end and a back-end, that exchange data
over an intermediate modeling language.

The functional and architectural specifications are first translated
into an intermediate model. The semantics of the intermediate language
is Stochastic Automata Networks and is described in details
in~\cite{stones_pinto_2010,MbdusPintoAllerton2010}. The functional
specification is bound to the hardware components at the level of the intermediate language (because function
and architecture are defined using different languages). The user
provides binding information as input to the tool. The intermediate
model is then passed to a back-end tool for analysis.

The first step in the analysis of the model is to compute the set of
states that can be reached by the system. In fact, the intermediate
model is in the form of a set of automata that interact using
synchronization primitives. This interaction restricts the set of
reachable states.  The intermediate model is first parsed and then
encoded into a Binary Decision Diagram~\cite{bdd,Somenzi98cudd:cu} to
perform symbolic reachability analysis. The result of the reachability
analysis is the set of all reachable states. It is possible to store,
as a byproduct of the reachability algorithm, the set of transitions
between reachable sates. This set can be used to construct the
infinitesimal generator of the Markov Chain (MC) underlying the
system. The MC is then solved by standard techniques for transient
analysis, or it can be used for probabilistic model
checking.  If
the goal is performance analysis, the tool allows the user to provide
a configuration file that can be used to filter the data and provide
projections of the results along some of the states
(e.g. ``probability of being in an unsafe state at all time''). The
tool also allows to define parameters instead of numeric transition
rates. These parameters can be used for quick comparison of different
system configuration, or they can be directly used in optimization
problems.

Remarkably, our implementation shows that the front-end
development effort ($6$ thousands lines of code) is comparable with
the back-end development effort, even for the restricted subset of the
input languages that we are able to translate at the moment. This
result highlights the importance of the model extraction problem, from
high-level description to analyzable probabilistic models.

\subsection{Examples}\label{sec:applications}

\paragraph{Example of functional analysis and synthesis}
\label{sec:an-auton-miss}
Consider an autonomous helicopter which is assigned the mission of
finding a building marked with a special symbol in a urban area.
Since the vision algorithm used to match the symbol against a known
pattern is sensitive to scaling, the position estimation error (caused
by the finite accuracy of the GPS and other sensors) can cause either
a false negative (i.e. the symbol is missed), or a false positive
(i.e. an object is recognized as the symbol). Each object with a
minimum level of matching is kept in a table with an associated
score. At each frame the score is updated depending on the quality of
the matching. We discretized the score into three levels: good ($g$),
average ($a$) and bad ($b$). We assume that there are four objects randomly
placed in the scene and that object $0$ is the symbol.

\begin{figure}[t!]
\centering
\includegraphics[width=0.70\hsize]{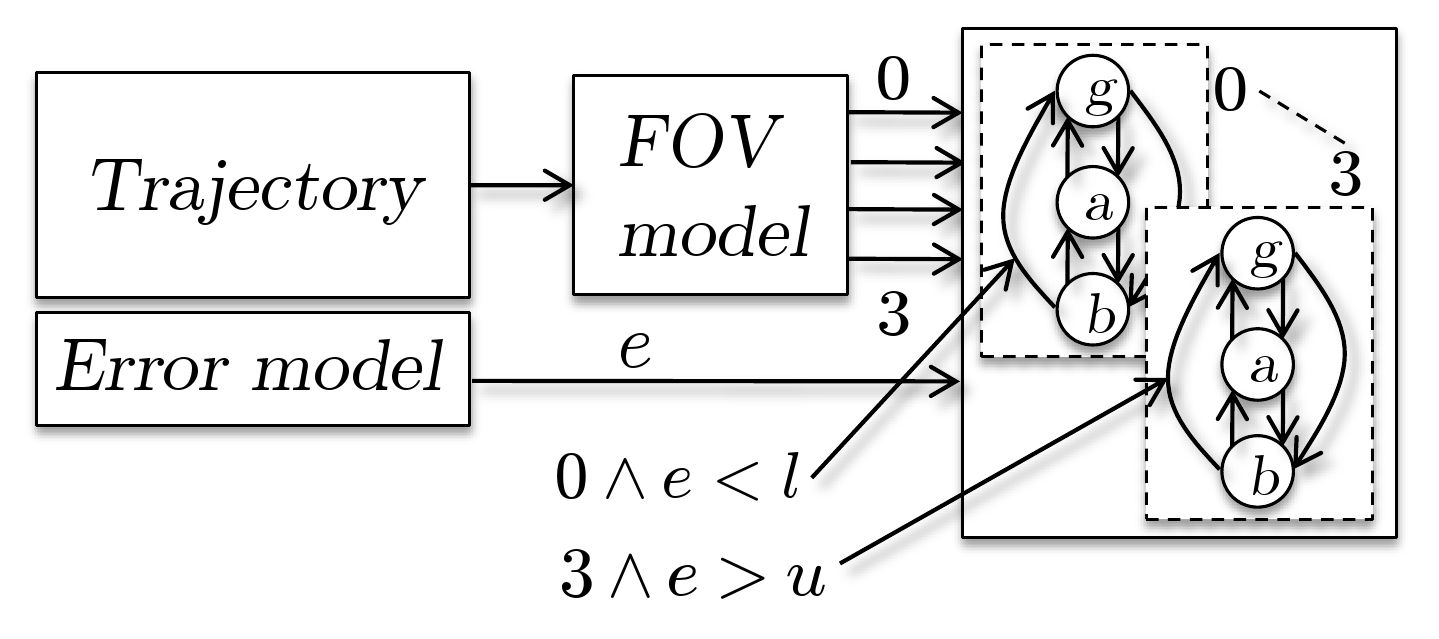}
\caption{High level diagram of the model of the autonomous mission.}
\label{fig:experiment_setup}
\end{figure}

The model is shown in Figure~\ref{fig:experiment_setup}\footnote{A
more detailed Simulink/Stateflow model can be found
  in~\cite{stones_case}} and has two parts. The trajectory followed by
the helicopter is computed by a trajectory generation algorithm for
given way-points around the building.  A camera model is used to
generate Boolean flags that are equal to TRUE if the corresponding
object is in the field of view of the camera and FALSE otherwise ({\tt
  Field of view model}).  The vision algorithm is a finite state model
that maintains a matching score for each of the objects in the scene
(the object being parallel states).  When in the field of view, the
score assigned to object $0$ increases if the position error is below
a lower bound $l$, and decreases if it is above an upper bound $u$
(transitions are reversed for the other objects). Because the update
is done at each frame, higher computation rates improve the ability to
distinguish image features. The system is translated into a SAN, where
the variance $\sigma$ of the colored noise $e$ is one of the
parameters. The transition rate $\lambda$ associated with transitions
in the vision algorithm model is another parameter. These two
parameters represent the accuracy of the sensors and the speed of
execution (i.e. frames per second) of the vision algorithm,
respectively. The helicopter follows a circular trajectory around the building. We
report the values of the probability of being in each of the three
level of the matching score for all the objects in
Table~\ref{tab:functional}.
%For very accurate sensors ($\sigma=0.21$),
%low computation rates can still guarantee a resonably high probability
%of successfully distinguishing the symbol. For noisy sensors, even
%high computation rates cannot guarantee an acceptable probability of
\begin{table}
  \centering
{\footnotesize
  \begin{tabular}{cccccc}
    $(\lambda,\sigma)$ & Score &  0 &  1 &  2 &  3 \\
    \hline
    & $g$ & 0.6207 &   0.0254 & 0.0110 & 0.0241 \\
    $(0.5,0.21)$ & $a$ & 0.1846 & 0.1823 & 0.1792 & 0.1805 \\
    & $b$ & 0.1946   & 0.7923 & 0.8098 & 0.7954 \\
    & $g$ & 0.5561   & 0.0472 & 0.0381 & 0.0456 \\
    $(1,0.42)$ & $a$   & 0.3821 & 0.3741 & 0.3806 & 0.3738 \\
    & $b$ & 0.0617   & 0.5787 & 0.5813 & 0.5806 \\
    & $g$ & 0.4071   & 0.1424 & 0.1401 & 0.1423 \\
    $(10,0.84)$ & $a$  & 0.4533 & 0.4527 & 0.4526 & 0.4529 \\
    & $b$ & 0.1396   & 0.4048 & 0.4073 & 0.4047 \\
    \hline
  \end{tabular}
}
  \caption{Probabilities of good, average and bad matching for different values of error variance and computation speed.}
  \label{tab:functional}
\end{table}

Recently we have also addressed the synthesis problem of designing a control strategy for a vehicle that provides probabilistic guarantees of accomplishing the mission in a hostile environment with obstacles and moving adversaries \cite{synthesis}. The vehicle is required to satisfy a mission objective expressed as a temporal logic specification over a set of properties satisfied at regions of a partitioned environment.  We capture the motion of the vehicle and the vehicle updates of adversaries distributions as a Markov Decision Process. In particular, we abstract the probability of a vehicle computing a feasible path and avoiding obstacles over a region under assumption that the trajectory planner is using a sampling-based motion planning algorithm such as one described in Section~\ref{sec:algorithm}. Using tools in Probabilistic Computational Tree Logic, we find a control strategy for the vehicle that maximizes the probability of accomplishing the mission objective. For details reader is referred to \cite{synthesis}.

\paragraph{Example of architectural analysis}
\label{sec:arch-analys}
We consider a distributed architecture composed of processors running
a single thread and communicating over a token ring bus (details can be found
in~\cite{MbdusPintoAllerton2010}). Each thread $th_i$ has three states: a
$sleep$ state where the thread is not active, a $ready$ state where the thread
is ready to be executed by the processor is busy, and a $run$ state where the
thread is running. The thread transitions from the $sleep$ state to the $ready$
state when its transmission buffer $TX_i$ is empty. When the thread is
scheduled to run, it first reads from buffer $RX_i$, and then writes its
transmission buffer $TX_i$. A token ring bus serves the TX buffers and
broadcasts their content to all RX buffers in the system. We consider
transition rates of $10^5$, $10^4$ and $10^3$ for transitions $(sleep,ready)$,
$(ready,run)$ and $(run,sleep)$ respectively. We also consider a rate of $8000$
for the protocol to pass the token among users, while we leave the data
transmission rate $\lambda$ as a parameter (to mimic the effect of different
packet sizes). We consider three architectures: $sys2$ with two processors
($182$ reachable states), $sys4$ with four processors ($24708$ reachable
states) and $sys4f$ with 4 unreliable processors ($2118680$ reachable states).
Unreliable processors can fail with rate $0.0003$, and recover from failure
with rate $0.3$. The results of the analysis are shown in
Table~\ref{tab:architectural} where we report the probability of being in the
sleep, ready or run state for thread $th_2$ at time $t = 1ms$.

\begin{table}
  \centering
{\footnotesize
  \begin{tabular}{ccccc}
    $\lambda$ & System & $P(sleep)$ & $P(ready)$  & $P(run)$ \\
    \hline
    & $sys2$ & 0.275  & 0.058 & 0.666 \\
    8000 & $sys4$ & 0.380 & 0.038 & 0.581 \\
    & $sys4f$ & 0.371 & 0.038 & 0.591 \\
    & $sys2$ & 0.378& 0.040 & 0.581 \\
    4000 & $sys4$ & 0.453 & 0.025 & 0.522 \\
    & $sys4f$ & 0.439 & 0.025 & 0.535 \\
    & $sys2$ & 0.459 & 0.026 & 0.515 \\
    2000 & $sys4$ & 0.505 & 0.016 & 0.479 \\
    & $sys4f$ & 0.489 & 0.016 & 0.495 \\
    \hline
  \end{tabular}
}
  \caption{Probabilities of being in the sleep, ready or run state at $t=1ms$ for thread $th_2$.}
  \label{tab:architectural}
\end{table}
The results show two obvious trends. When the number of processors
increases, the token rotation time increases and the time a task
spends in the sleep state also increases. If the transmission time
increases, the time spent in the sleep state also
increases. Interestingly, the time spent in the run state is higher
for $sys4f$ than for $sys4$. This is because thread $th_2$ can
leverage the time when other processors are silent because of a
failure.

\section{Conclusions}
In this paper we have overviewed some recent advances in methodology and tools to model, analyze, and design
interconnected dynamical systems with emphasis on robust autonomous aerospace systems operating in uncertain environment. The key idea that enables scalable computation and analysis is the decomposition of the large interconnected systems into weakly interacting subnetworks. We reviewed a new  graph decomposition method based on wave equation based clustering to identify such weakly interacting subnetworks in a scalable manner. We showed how this decomposition can be exploited to accelerate uncertainty quantification in the probabilistic waveform relaxation approach. We also addressed another aspect of uncertainty management: design of search and tracking algorithms for unmanned aerospace systems that allow optimal uncertainty reduction in location of stationary and mobile targets in search and tracking problems. We described a novel ergodic dynamical system theory based sensor resource management approach for such applications. Robust motion planning using probabilistic roadmap was also demonstrated for realistic vehicle dynamics in an obstacle rich environment. Building on these approaches which apply to a specific abstraction level, we presented a general unified abstraction-based methodology and tools for analysis and synthesis for uncertain systems. We demonstrated the efficient uncertainty quantification and robust design using the case studies of model-based target tracking and search, and mission planning in an obstacle rich environment. To show the generality of our methodology we also considered problem of uncertainty quantification in energy usage in buildings, and stability assessment of interconnected power networks. We also briefly identified several challenges and promising future directions for each methodology and tool.

\section{Acknowledgement}
This work was supported in part by DARPA DSO under AFOSR contract FA9550-07-C-0024, Robust Uncertainty Management (RUM). Any opinions, findings and conclusions or recommendations expressed in this material are those of the author(s) and do not necessarily reflect the views of the AFOSR or DARPA. The authors will like to thank Jerrold E. Marsden, Matthew West, Sean P. Meyn, Calin Belta, Michael Dellnitz, Sudha Krishnamurthy, Suresh Kannan,  Konda Reddy Chevva, Sanjay Bajekal, Sophie Lorie and  Jose M. Pasini, for useful discussions and feedback. We would especially like to acknowledge the profound influence that Jerry Marsden had on this research direction and on many of us personally. Jerry brought to us the idea of using graph decomposition to break the complexity of computations and analysis in uncertain dynamic networks, the key idea behind the RUM project. In fact, Jerry was the person who invented the acronym DyNARUM (Dynamic Network Analysis for Robust Uncertainty Management), which was  how we internally called the RUM project. Jerry Marsden was also one of the champions of using precomputed piecewise optimal motion primitives.  Finally, Jerry Marsden has personally inspired many of us to pursue some the key research ideas included in this paper. We called the positive influence of the discussions with Jerry Marsden that stayed with us for a long time after the discussions and kept us pursuing the research with hope and excitement ``the Jerry effect". This paper is a result of the ``Jerry effect" that is still within us.

\bibliographystyle{plain} %elsarticle-harv
\bibliography{pwrbib}

\end{document}